\documentclass[epj,nopacs]{svjour}
\usepackage{amssymb}
\usepackage[dvips]{graphicx}
\usepackage{amsmath}
\usepackage{bm}
\usepackage{epsfig}
\usepackage{widetext}

\usepackage{atbegshi}% http://ctan.org/pkg/atbegshi
\AtBeginDocument{\AtBeginShipoutNext{\AtBeginShipoutDiscard}}
\begin{document}

%%%%%%%%%%%%%%%%%%%%%%%%%%%%%%%%%%%%%%%%%%%%%%%%%%%%%%%%%%%%%%%%%%%%%%%%%%%%%%%%%%%%%%%%%%%%%%%%%%%%%%%%%%%%%%%%%%%%%%%%%%%%

\title{Higher-order gravity in higher dimensions: Geometrical origins of four-dimensional cosmology?}
\author{Antonio Troisi\inst{1}}
%\email{atroisi@unisa.it}
\thanks{{e-mail:}\emph{atroisi@unisa.it}}
%\affiliation{Dipartimento di Fisica ``E.R. Caianiello", Universit\`a degli Studi di Salerno, Via Giovanni Paolo II, 132, 84084, %Salerno, Italy.}
\institute{Dipartimento di Fisica ``E.R. Caianiello", Universit\`a degli Studi di Salerno, Via Giovanni Paolo II, 132, 84084, Salerno, Italy.}
\abstract{Determining cosmological field equations represents a still very
debated matter and implies a wide discussion around different
theoretical proposals.
A suitable conceptual scheme could be represented by gravity models
that naturally generalize Einstein
Theory like higher order gravity theories and higher dimensional ones.
 Both of these two different approaches allow to define, at the effective level, Einstein field equations equipped with source-like energy momentum tensors of
geometrical origin. In this paper, it is
discussed the possibility to develop a five dimensional fourth order gravity
model whose lower dimensional reduction could provide an interpretation of
cosmological four dimensional matter-energy components. 
We describe the basic concepts of the model, the complete field equations formalism and the 5-Dim to 4-Dim reduction procedure. Five dimensional $f(R)$ field equations turn out to be equivalent, on the four dimensional hypersurfaces orthogonal to the extra-coordinate, to an Einstein like cosmological model with three matter-energy tensors related with higher derivative  and higher dimensional counter-terms. By considering a gravity model $f(R)=f_0R^n$ it is investigated the possibility to obtain five dimensional power law solutions. The effective four dimensional picture and the behaviour of the geometrically induced sources are finally outlined in correspondence to simple cases of such higher dimensional solutions.}
\date{\today}

%\pacs{04.50.Kd, 04.50.Cd, 11.10.Kk, 98.80.-k}

\maketitle
%%%%%%%%%%%%%%%%%%%%%%%%%%%%%%%%%%%%%%%%%%%%%%%%%%%%%%%%%%%%%%%%%%%%%%%%%%%%%%%%%%%%%%%%%%%%%%%%%%%%%%%%%%%%%%%%%%%%%%%%%%%%

%%%%%%%%%%%%%%%%%%%%%%%%%%%%%%%%%%%%%%%%%%%%%%%%%%%%%
\section{Introduction}
%%%%%%%%%%%%%%%%%%%%%%%%%%%%%%%%%%%%%%%%%%%%%%%%%%%%

Type Ia supernovae (SNeIa) observations depicted a late-time speeding up universe \cite{RiessPerlmutter,SNeIa-2,SNeIa-3}, driven by an unknown  component, whose properties can be ascribed to some sort of exotic fluid. This elusive component has been addressed to a form of dark energy that exhibits an anti-gravitational negative equation of state (EoS) \cite{sahnipeebles}. The combination of standard matter and dark energy provides an Einstein-like Universe where gravitational attraction is counterbalanced by such an unusual gravitational source therefore reproducing cosmological observations. Soon after this discover a plethora of theoretical proposals has been suggested in order to explain dark energy origin. However, at today, a well endowed and self consistent physical interpretation is so far unknown \cite{burgess2013,sami2006,DEreview}. The puzzling quest of a satisfactory explanation about these phenomenological results led cosmologists to explore several research lines. In particular, people followed two main directions. From one side, standard Einstein gravity has been reviewed by introducing a new cosmological component: the cosmological constant or a whatsoever well behaved fluid with a negative EoS. Differently, adopting an unconventional point of view, they have been considered modified gravity models that generalize Einstein theory: i.e. scalar-tensor gravity \cite{Brans1961,Bartolo1999,FaraoniST}, $f(R)$ theories \cite{curv1,CapozFar,NojOd}, DGP gravity \cite{Dvali2000}, braneworld scenarios \cite{brax2003,MaartensLR,RS}, induced matter theory \cite{wessbook1,KK,wessbook2} and so on.

Among others $f(R)$ theories of gravity received a considerable attention.  Such theories, which represent a natural generalization of Einstein gravity, are obtained by relaxing the hypothesis of linearity contained in the Hilbert-Einstein Lagrangian. Several results, sometimes controversial, have been obtained in this framework \cite{f(R)Review}. These models have been satisfactory checked with cosmological observations \cite{curvmimick,husaw,starobL} and suggest intriguing peculiarities in the low energy and small velocity limit \cite{curvlow} since the gravitational potential displays a Yukawa like correction \cite{stelle}. In order to be viable $f(R)$ gravity theories have to satisfy some minimal prescriptions. In particular, they have to match cosmological observations avoiding instabilities and ghost-like solutions and they have to evade solar system tests \cite{defeliceRev}. Constraints from energy conditions represent a further argument to settle a suitable gravity Lagrangian \cite{alcaniz2007}. However, at this stage, a fully satisfactory fourth order gravity model is far to be achieved. As a matter of fact, such kind of models have been considered also from other points of view. For example, recently, models with a non-minimal coupling between gravity and the matter sector have been taken into account \cite{nonminRmatt,fRT}.  

On the other side, historically, attempts for a unification of gravity with other interactions stimulated the search for theoretical schemes based on higher dimensions, i.e. beyond our conventional four dimensional (4-D) spacetime. Nordstr{\o}m \cite{nordstrom1914}, who was the first to formulate a unified theory based on extra dimensions, Kaluza \cite{kaluza1921} and Klein \cite{klein1926} developed a five-dimensional (5-D) version of General Relativity (GR) in which electrodynamics derivates as a counter-effect of the extra-dimension. Successively, a lot of work has been dedicated to such theoretical proposal both considering  fifth-coordinate compactification and large extra-dimensions, allowing for non compact mechanisms \cite{KK}. This last approach led to the so called  Space-Time-Matter (STM) or Induced-Matter  Theories (IMT) \cite{wessbook1,wessbook2} and to braneworld theories \cite{brax2003}. The basic characteristic of IMT is that 5-D vacuum field equations can be recast after the reduction procedure as 4-D Einstenian field equations with a source of geometrical origin. In such an approach, the matter-energy source of 4-D spacetime represents a manifestation of extra dimensions. Generalizations of Kaluza-Klein theory represent themselves a suitable scheme to frame modern cosmological observations. In fact, after SNIe observations, several attempts have been pursued to match Induced Matter Theories with the dark energy puzzle  \cite{aguilar2008,PdeLeon2010,deLeonjcap2010,rasoulietal}. 

In this paper we try to merge the two approaches. Since vacuum fourth order gravity theories, as well as IMT, can be cast as an Einstein model with a matter-energy source is of geometrical origin  \cite{curv2,curv3}, it seems a significant proposal to confront the two theoretical schemes. The genuine idea underlying this work is to investigate 5-D $f(R)$-gravity from the point of view of a completely geometric self-consistent approach. In this scheme, all matter-energy sources will represent a byproduct of the dimensional reduction related to higher order terms and higher dimensional quantities. Such an approach determines a completely different conceptual framework with respect to standard five dimensional fourth order gravity models. The main purpose is to recover along this scheme both dark matter and dark energy dynamical effects.  
In the standard realm, in fact, extra dimensions and curvature counter-terms will only play the role of mending dark energy phenomenology in presence of ordinary matter. Along this orthodox paradigm, 5-D models of fourth order gravity have been studied in presence of perfect fluid sources \cite{vajdi2009,huang2010}  under peculiar assumptions on the metric potentials. In a similar fashion, accelerating 4-D cosmologies, induced by generalized 5-D $f(R)$ gravity, have been also studied considering a curvature-matter coupling \cite{wu2014}. Furthermore, a vacuum fourth order Kaluza-Klein theory, defined in term of the Gauss-Bonnet invariant, has been studied from the point of view field equations predictions. In particular,  it has been investigated, in the cylinder approximation, the propagation of its electromagnetic degrees of freedom \cite{schimming2002} and the particle spectrum in the linear regime \cite{schimming2003}.  
\\
Attempts to meet $f(R)$-gravity and IMT have been also developed in time. For example it has been investigated the possibility to draw information on the Space-Time-Matter tensor starting from the energy-momentum tensor induced by higher order curvature counter-terms \cite{darabi2009}. On the other side a new effective coupled $F(^{(4)}\! R,\varphi)$ gravity theory has been proposed as a consequence of a five dimensional $f(R)$ model \cite{aguilar2015}. Within such a paper it has been also demonstrated that the Dolgov-Kawasaki stability criterion of  the 5-D $f(R)$ theory remains the same as in usual $f(R)$ theories: $f''(R) > 0$, with prime that indicates the derivative with respect to the Ricci scalar. 
\\
In our work we try to develop a more general framework, at first 5-D $f(R)$-gravity is discussed in a complete analytical form. We provide general field equations in the $f(R)$-IMT approach and, by exploiting the reduction procedure from 5-D to the 4-D ordinary spacetime (considering a suitable extension of Friedmann-Robertson-Walker metric), we obtain a complete set of new Einstein-like field equations with matter-energy sources induced by higher order derivative terms and higher dimensions (top-down reduction).  In other words, we develop a fully geometrical cosmology where ordinary matter and dark components could be addressed, in principle, as the outgrowth of the GR formalism adopted within the top-down reduction mechanism. 
\\
The paper is organized as follows. In section II we provide a brief review of 4-D $f(R)$-gravity. Section III is devoted to summarize Induced Matter Theory. The 5-D $f(R)$-gravity formalism is outlined in section IV, and, thereafter, we discuss the 5-D to 4-D reduction procedure of field equations. In section V  it is described a routine to find 5-D power law solutions. Section VI is dedicated to the analysis of the results and, in particular, to their interpretation from the point of view of the 4-D induced cosmologies. Finally, Section VII is left to conclusions.

%%%%%%%%%%%%%%%%%%%%%%%%%%%%%%%%%%%%%%%%%%%
\section{The Cosmological Equations for $f(R)$  Gravity} \label{sectionfR}
%%%%%%%%%%%%%%%%%%%%%%%%%%%%%%%%%%%%%%%%%%%

Dark energy models are based on the
underlying assumption that Einstein's General Relativity is indeed
the correct theory of gravity. However, adopting a different point of view,  both cosmic speed up and dark matter can be addressed to a 
breakdown of GR. As a matter of fact, 
one should consider the possibility to generalize the Hilbert-Einstein (H-E)
Lagrangian. With these premises in mind, the choice of the gravity Lagrangian can be settled by means of data with the only prescription of adopting an ``economic" strategy, i.e. only minimal generalization of H-E action are taken into account. The so called $f(R)$-gravity \cite{curv1,f(R)Review,buchdal}, which consider an analytic function in term of Ricci scalar and provides fourth order field equations, is based on this conceptual scheme. It has to be reminded that higher order gravity theories represent the natural effective result of several theoretical schemes \cite{oneloop}, i.e. quantum field theories on curved spacetimes and M-theory. In addition, they have been widely studied as inflationary models in the early universe \cite{starobinsky,kerner}. 

Fourth order gravity is favored by Ostrogradski
theorem. It has been in fact demonstrated \cite{woodard} that $f(R)$-Lagrangians are 
the only metric-based, local and potentially stable modifications of gravity among the several
that can be constructed by means of the curvature tensor and possibly by means of 
its covariant derivatives. 

Let us consider the generic $f(R)$ action, we have: 

\begin{equation} \label{f(R)action}
S=\frac{1}{2}\int d^4 x \sqrt{-g} \, \left[f(R) 
+S_M(g_{\mu\nu},\psi)\right],
\end{equation}
where we have used natural units $8\pi G=c=\hbar =1$,  
$g$ is the determinant of the metric and $R$ is the Ricci scalar 
 and $\psi$ characterizes the matter fields.
Varying with respect to the metric we get field equations

 \begin{equation}\label{efef(R)}
 f'(R)R_{\mu\nu}-\frac{1}{2}f(R)g_{\mu\nu}- 
\left[\nabla_\mu\nabla_\nu -g_{\mu\nu}\Box\right] f'(R)= 
 \,T^M_{\mu\nu}, 
\end{equation}
where, the prime, as said in the introduction, denotes derivative with respect to $R$ and, as usual, 

\begin{equation}
\label{set}
T^{M}_{\mu\nu}=\frac{-2}{\sqrt{-g}}\, \frac{\delta
S_M }{\delta g^{\mu\nu} }.
\end{equation}
Tracing (\ref{efef(R)}) we obtain

\begin{equation}\label{efef(R)T}
 f'(R)R-2f(R) + 3\Box f'(R)  = \,T^{M},
\end{equation}
where $T^{M}=g^{\mu\nu}T^{M}_{\mu\nu}$, so that the relation between the Ricci scalar and $T^{M}$ is obtained by means a differential equation, differently than GR where  $R=- \,T^M$. This result suggest that$ f(R)$-gravity field equations admit a larger variety of solutions than 
Einstein's theory. In particular, $T^{M}=0$ solutions will no longer
imply Ricci flat cosmologies and therefore that $R=0$. Actually, a suitable property of this model is that field equations (\ref{efef(R)}) can be recast in the Einstein form \cite{curv1,curv2}:

\begin{equation}\label{efef(R)E}
G_{\mu\nu} = R_{\mu\nu}-\frac{1}{2}g_{\mu\nu}R =
T^{curv}_{\mu\nu}+T^{M}_{\mu\nu}/f^\prime(R),
\end{equation}
with

\begin{displaymath}
T^{curv}_{\mu\nu}\,=\,\frac{1}{f'(R)}\Big\{\frac{1}{2}g_{\mu\nu}\left[f(R)-Rf'(R)\right]
+
\end{displaymath}
\begin{equation} \label{6}
f'(R)^{;\alpha\beta}(g_{\mu\alpha}g_{\nu\beta}-g_{\mu\nu}g_{\alpha\beta})
\Big\}\,,
\end{equation}
that represents a {\it curvature stress\,-\,energy tensor} induced by higher order derivative terms (the terms $f^\prime(R)_{;\mu\nu}$ render the equations of fourth order). The limit $f(R) \rightarrow R$ reduces Eqs.(\ref{efef(R)E}) to the
standard second-order Einstein field equations. 

The Einstein-like form of fourth order field equations (\ref{efef(R)E}) suggests that higher order counter-terms can play the role of a source-like component within gravity field equations. In practice, it is possible to postulate that geometry can play the role of a mass-energy component when higher than second order quantities are taken into account. Indeed, the trace equation (\ref{efef(R)T}) propagates a scalar-like degree of freedom. As a matter of fact, in principle, one can imagine to address Universe dark phenomenology to such a kind of effective fluid. From the cosmological point of view a relevant role is played by the barotropic factor of the curvature matter-energy fluid. In the vacuum case this quantity will discriminate accelerating solutions from standard matter ones. In the case of a Friedman spacetime, for example, one has \cite{curv2,curv3}: 

\begin{equation}
\rho_{curv} = \frac{1}{f'(R)} \left \{ \frac{1}{2} \left [ f(R)  -
R f'(R) \right ] - 3 H \dot{R} f''(R) \right \} \ , \label{eq:
rhocurv}
\end{equation}
\begin{equation}
w_{curv} = -1 + \frac{\ddot{R} f''(R) + \dot{R} \left [ \dot{R}
f'''(R) - H f''(R) \right ]} {\left [ f(R) - R f'(R) \right ]/2 -
3 H \dot{R} f''(R)} \ , \label{eq: wcurv}
\end{equation}
where $H=\dot{a}/a$ stands for the Hubble parameter and dot means a time derivative. 

In the following we will resort to a similar quantity in order to check the behaviour of the cosmological fluids induced by 5-D $f(R)$-gravity on the 4-D hypersurfaces that slice the higher dimensional spacetime along the extra-coordinate.
\\
In our conventions we will consider latin indices like $A,B,C,$ etc. running from $0$ to $4$, latin indices like $i,j,$ etc., running from $1$ to $3$ and greek indices taking values from $0$ to $3$.
%

%%%%%%%%%%%%%%%%%%%%%%%%%%%%%%%%%%%%%%%%%%%%%%%%%%
\section{Induced Matter Theory}\label{secIMT}
%%%%%%%%%%%%%%%%%%%%%%%%%%%%%%%%%%%%%%%%%%%%%%%%%%

Modeling a unification theory that contains gravity and particle physics forces, typically, implies the resort to extra-dimensional models \cite{KK,carlip}. Among these 5-D Kaluza-Klein theory \cite{kaluza1921,klein1926} and its modern revisitations Induced-Matter and membrane theory represent significant approaches; in addition these kind of models represent the low energy limit of more sophisticate theories (i.e. supergravity) 	\cite{kaku,suNgra}. The main difference between the seminal approach of Kaluza and Klein and Induced Matter Theory is related with the role of the extra-dimension. In its first conception the fifth dimension was ``rolled up'' to a very small
size, answering the question of why we do not ``see'' the fifth dimension. Modern theories like IMT postulate that we are constrained to live in a smaller 4-D hypersurface embedded in a higher dimensional spacetime. The key-point for embedding the 4-D Einstein theory into the 5-D Kaluza-Klein Induced-Matter-Theory is represented by Campbell-Magaard (CM) theorem. The problem of embedding a Riemannian manifold in a Ricci-flat space was studied by Campbell soon after GR discovery \cite{cambell1926}, and finally demonstrated by Magaard in 1963 \cite{magaard1963}. Further on, Tavakol and coworkers \cite{romero96} used these studies to establish mathematical well endowed bases to the 4-D reinterpretation of 5-D Kaluza-Klein theory that is dubbed Induced-Matter Theory \cite{wesson1992,wessPdL92}. In such a framework, the 5-D to 4-D reduction procedure determines on the 4-D hypersurface an Einstein gravity theory plus induced matter components of geometrical origin. The CM theorem can be formalized as follows: 

\vspace{5mm}
{\it Any analytic Riemannian space $V_{n}\left( x^\mu,t\right) $ can be locally embedded in a Ricci-flat Riemannian
space $V_{n+1}\left( x^A,t\right) $.}
\vspace{5mm}

Here the ``smaller'' space has dimensionality $n$ with $\mu=0,..., n-1$, while the ``host'' space has
dimensionality $N=n+1$ with $A$ running from 0 to $n$, the extra-coordinate can be both space-like and time-like. We are interested to the case $N=5$. It is important to remark that CM theorem is a  {\it local}
embedding theorem. Therefore, more general issues related with global embedding, i.e. initial-value problems, stability or general induced solutions \cite{kernerEmb}, cannot be resort to this achievement. However, for our purposes, the theorem guarantees the right analytic framework in order to frame 4-D matter phenomenology in relation to 5-D field equations \cite{wessPdL92}. 

Because of CM theorem it is possible to write down the 5-D metric  as follows:

\begin{equation}\label{5Dmetric}
dS^2 = \gamma_{AB}d x^A d x^B = g_{\mu\nu}(x, y)d x^{\mu}d x^{\nu} + \epsilon \Phi^2(x, y) d y^2,
\end{equation}
where $x^{\mu} = (x^0, x^1, x^2, x^3)$ are 4-D coordinates, $g_{\mu\nu}$ turns out to be the spacetime metric and $y$ is the fifth coordinate, $\epsilon = \pm 1$ allows for space-like or time-like extra dimension. Along the paper we adopt a spacetime signature $(+, -, -, -)$. 
With these premises in mind ordinary $4D$ spacetime result as a hypersurface $\Sigma_{y}:y =  y_{0} = $ constant, orthogonal to the $5D$ extra-coordinate basis vector

\begin{equation}
\label{unit vector n}
{\hat{n}}^{A} = \frac{\delta^{A}_{4}}{\Phi}, \;\;\;n_{A}n^{A} = \epsilon.
\end{equation}
If one considers vacuum 5-D field equations (Ricci flat): $R_{AB}=0$, the CM theorem, suggests a natural reduction process to a 4-D pseudo-Riemannian spacetime. In practice, one can obtain an Einstein-like 4-D model\footnote{We recall that natural units are adopted, therefore $8\pi G\diagup c^{4}$ is set equal to unity.} $G_{\mu\nu}=T_{\mu\nu}$ \cite{wesson1992,wessPdL92}, once the right hand side of these equations fulfills the definition 

\begin{align}\label{IMT}
T_{\mu\nu}^{IMT}&=\frac{\Phi_{,\mu;\nu}}{\Phi}-\frac{\epsilon}{2\Phi^2}\left\{\frac{\stackrel{\ast}\Phi \stackrel{\ast}g_{\mu\nu}}{\Phi} - \stackrel{\ast \ast}g_{\mu\nu}+g^{\alpha\beta}\stackrel{\ast}g_{\mu\alpha}\stackrel{\ast}g_{\nu\beta}+\right.\nonumber
\\
&\left.- \frac{g^{\alpha\beta}\stackrel{\ast}g_{\alpha\beta}\stackrel{\ast}g_{\mu\nu}}{2} + \frac{g_{\mu\nu}}{4}\left[\stackrel{\ast}g^{\alpha\beta}\stackrel{\ast}g_{\alpha\beta} + \left(g^{\alpha\beta}\stackrel{\ast}g_{\alpha\beta}\right)^2\right]\right\}.
\end{align}
Here, a comma is the ordinary partial derivative, a semicolon denotes the ordinary 4D covariant and starred quantities describe terms derived with respect to the fifth coordinate. This quantity defines the so called Induced Matter Tensor; the only hypothesis underlying this achievement has been  relaxing the cylinder condition within the Kaluza-Klein scheme (independence on the fifth coordinate).

As it is possible to observe from the previous result, fourth order gravity theories and Kaluza-Klein IMT models provide the same conceptual scheme.  Both of the approaches allow to obtain an Einstein-like gravity model where matter-energy sources are of geometrical origin. In particular, these effective matter-energy tensors descend, respectively, from the higher order derivative contributions and from the higher dimensional counter-terms. Therefore, it seems that deviations from GR can be naturally recast  as sources of standard Einstenian models. \\
In that respect let us notice that IMT and 5-D $f(R)$ gravity are built in a different philosophy with respect to conventional higher dimensional approaches like braneworld models. In fact, despite the same working scenario: bulk universe with non trivial dependence on the extra-coordinate, 4-D metric obtained evaluating the background metric at specific 4-D hypersurfaces, matter fields confined in the 4-D spacetime, there are intrinsic conceptual differences. 5-D $f(R)$ gravity and Induced Matter Theory are based on the hypothesis that standard matter is nothing else than a 4-D manifestation of geometrical deviations with respect to GR. Braneworlds model our universe as a four dimensional singular hypersurface, the brane, embedded in a five dimensional anti De Sitter spacetime. In such a case the motivation for a non compactified extra-dimension is to solve the hierarchy problem. The differences in term of physical motivations for large extra-dimensions imply also different technical approaches. Within IMT and 5-D $f(R)$ gravity one considers a Ricci flat vacuum bulk and develops the 4-D physics as a byproduct. In braneworlds it is assumed the opposite point of view. In this case, one deals with suitable 4-D solutions on the brane for some matter distributions and these solutions are matched with an appropriate 5-D bulk considering Israel junction conditions. Nevertheless, despite such conceptual differences it has been showed in time that brane theory, IMT and 5-D $f(R)$ gravity can have a suitable matching. In particular, IMT and braneworlds have been put in strict analogy \cite{PdLMPLA01}. The key point is to consider the induced matter approach in term of the spacetime extrinsic curvature. In this case it is possible to write down braneworld-like field equations with a 4-D matter-energy source and a brane tension that are defined in term of the extrinsic curvature. On the same line very recently it has been shown that $Z_2$ braneworld-like solutions can be obtained as particular maximally symmetric solutions of 5-D $f(R)$ gravity with matter \cite{aguilar2015}. In future works one can imagine to deepen the interconnections that seem to arise among these different higher dimensional approaches.\\ 
Such considerations suggest, in principle, the intriguing possibility that GR experimental shortcomings could be actually addressed to the effective property of modified gravity models. In the following we will exploit such approach to study a general fourth order 5-D model, where higher order derivative gravity is ``merged" with higher dimensions.

%%%%%%%%%%%%%%%%%%%%%%%%%%%%%%%%%%%%%%%%%%%%%%%%%%
\section{A 5-D $f(R)$-gravity model and its 4-D reduction}\label{f(R)5}
%%%%%%%%%%%%%%%%%%%%%%%%%%%%%%%%%%%%%%%%%%%%%%%%%%

A $f(R)$ theory of gravity in five dimensions can be described by the action
\begin{equation}\label{actionfR5}
{\cal S}^{(5)}=\frac{1}{2}\int d^{5}\!x\sqrt{g^{(5)}}f(R^{(5)})+\frac{1}{2}\int d^{5}\!x\sqrt{g^{(5)}}{\cal L}_{M}(g_{AB},\psi),
\end{equation}
where $R^{(5)}$ is the 5-D Ricci scalar, ${\cal L}_m(g_{AB},\psi)$ is a Lagrangian density for matter fields denoted, as in the 4-D case, by $\psi$ and $g^{(5)}$ represents the determinant of the 5-D metric tensor $g_{AB}$. 5-D field equations can be obtained by action (\ref{actionfR5}) varying with respect to the metric:
\begin{align}\label{equationsfR5}
 f_{,R}^{(5)}R^{(5)}_{AB}\, &- \,\frac{1}{2}\, f(R^{(5)})\, g_{AB}\,+\\\nonumber
& -\, [\nabla\!_A\nabla\!_B-g_{AB}\,\Box^{(5)}]\,f_{,R}^{(5)}= T^{(5)\ M}_{AB},
\end{align}
here $T^{(5)\ M}_{AB}$ is the energy-momentum tensor for matter sources, while $\nabla_A$ is the 5D covariant derivative, $\Box^{(5)} =g^{AB}\nabla_A\nabla_B$ is the 5D D'Alambertian operator. In order to simplify the notation we have defined $f_{,R}(R^{(5)})\equiv f_{,R}^{(5)}$, therefore henceforth with $f_{,R}^{(5)}$ it is intended the 5-D Ricci derivative of the $f(R^{(5)})$ gravity Lagrangian.\\

Together the field equations (\ref{equationsfR5}) one can obtain the trace 

\begin{equation}\label{p3}
4\,\Box^{(5)} f_{,R}^{(5)}+f_{,R}^{(5)}R^{(5)}-\frac{5}{2}f(R^{(5)})= T^{(5) M}
\end{equation}
where $T^{(5)M}= g^{AB}\,T^{(5)M}_{AB}$.\\
Actually, 5-D field equations (\ref{equationsfR5}) can be naturally recast as a generalization of Einstein 5-D equations in the same manner of the ordinary 4-D formalism given in section \ref{sectionfR}. In fact, with some algebra and isolating 5-D Einstein tensor one gets\,:

\begin{align}
& R^{(5)}_{AB}\!-\!\frac{1}{2}R^{(5)}g_{AB}\!=\!\frac{1}{f_{,R}^{(5)}}\left\{\frac{1}{2}\left(f(R^{(5)})\!-\!f_{,R}^{(5)}R^{(5)}\right)g_{AB}\right. \nonumber
\\\label{curvquifR5}
& \left.-\, [g_{AB}\,\Box^{(5)}-\nabla_A\nabla_B]\,f_{,R}^{(5)}\right\}\!+\!\frac{1}{f_{,R}^{(5)}}\,T^{(5)M}_{AB}\,,
\end{align}
by fact determining a 5-D new source of geometrical origin in the right member of field equations

\begin{align}
 T^{(5)\;Curv}_{AB}\!&=\!\frac{1}{f_{,R}^{(5)}}\left\{\frac{1}{2}\left(f(R^{(5)})-f_{,R}^{(5)}R^{(5)}\right)g_{AB}\!+\!
\right.\nonumber
\\\label{TcurvquifR5}
& \left.-\, [g_{AB}\,\Box^{(5)}-\nabla_A\nabla_B]\,f_{,R}^{(5)}\right\}\,.
\end{align}
This fact, which can resemble only the byproduct of mathematical trick, determines significant physical consequences at 5-D and, as a consequence, also in the ordinary spacetime. In fact, as we will see in the following, the 4-D dimensional reduction of Eqs.(\ref{curvquifR5}) and (\ref{TcurvquifR5}) implies a ``new'' set of Einstein-like equations with three matter-energy components all of geometrical origin. One of these quantities descends from the 4-D reduction of the Einstein tensor and the two terms derivate from the relative 4-D reduction applied to $T^{(5)\ Curv}_{AB}$. Since we want to explore how geometric counter-terms can effectively mimic cosmological sources we neglect ordinary matter, i.e. here on we will assume $T_{AB}^{(5)\ M}=0$, considering gravity equation in vacuum. 

Let us now develop the 5-D to 4-D reduction procedure of our higher dimensional framework. At first, by assuming the metric (\ref{5Dmetric}), it is possible to draw the reduction rules for the differential operators \cite{deLeonjcap2010}\,:

\begin{align}
\nabla_{\mu}\nabla_{\nu}f_{,R}^{(5)} &= D_{\mu}D_{\nu}f_{,R}^{(5)} + \frac{\epsilon}{2 \Phi^2}\stackrel{\ast}g_{\mu\nu}\stackrel{\ast}{f_{,R}^{(5)}},
\\
\nabla_{4}\nabla_{4}f_{,R}^{(5)} &= \epsilon \Phi \left(D_{\alpha}\Phi\right)\left(D^{\alpha}f_{,R}^{(5)}\right) + \stackrel{\ast \ast}{f_{,R}^{(5)}} -  \frac{\stackrel{\ast}\Phi}{\Phi}\stackrel{\ast}{f_{,R}^{(5)}}, 
\\
 \Box^{(5)}f_{,R}^{(5)} &= \Box f_{,R}^{(5)} + \frac{\left(D_{\alpha}\Phi\right)\left(D^{\alpha}f_{,R}^{(5)}\right)}{\Phi} +\nonumber
\\+& \frac{\epsilon}{\Phi^2}\left[\stackrel{\ast \ast}{f_{,R}^{(5)}} + \stackrel{\ast}{f_{,R}^{(5)}} \left(\frac{g^{\alpha\beta}\stackrel{\ast}g_{\alpha\beta}}{2} - \frac{\stackrel{\ast}\Phi}{\Phi}\right)\right]\,;
\end{align}
here, again, the asterisk denotes partial derivative with respect to the extra coordinate (i.e., $\partial/\partial y = ^\ast$);  $D_{\alpha}$ is the four-dimensional covariant derivative defined on the hypersurface $\Sigma_{y=y_0}$, calculated with $g_{\mu\nu}$, and the usual D'Alambertian $\Box$ is referred to 4-D quantities. In the same shape, all the quantities that are not labelled in term of the $5th$ coordinate will be intended to refer to ordinary spacetime. On this basis, it is possible to rewrite the 5-D field equations (\ref{curvquifR5}) by separating the spacetime part ($\mu=0,..,3;\ \nu=0,..,3$) from the extra-coordinate ($y$) one\,:
\begin{widetext}
\begin{align}\label{curvquifRspazio}
& G_{\mu\nu}^{(5)}=\!\frac{1}{f_{,R}^{(5)}}\left\{\frac{1}{2}\left(f(R^{(5)})\!-\!f_{,R}^{(5)}R^{(5)}\right)g_{\mu\nu}
-\left[\Box f_{,R}^{(5)} + \frac{\left(D_{\alpha}\Phi\right)\left(D^{\alpha}f_{,R}^{(5)}\right)}{\Phi} +\right.\right. \nonumber 
\\ 
&\left.\left.+  \frac{\epsilon}{\Phi^2}\left(\stackrel{\ast \ast}{f_{,R}^{(5)}} + \stackrel{\ast}{f_{,R}^{(5)}} \left(\frac{g^{\alpha\beta}\stackrel{\ast}g_{\alpha\beta}}{2} - \frac{\stackrel{\ast}\Phi}{\Phi}\right)\right) \right]g_{\mu\nu}
+\nabla_\mu\nabla_\nu\,f_{,R}^{(5)}+\frac{\epsilon}{2 \Phi^2}\stackrel{\ast}{g_{\mu\nu}}\stackrel{\ast}{f_{,R}^{(5)}}\right\}\!,
\end{align}

\begin{align}\label{curvquifRextra}
& G_{44}^{(5)}=\!\frac{1}{f_{,R}^{(5)}}\left\{\frac{1}{2}\left(f(R^{(5)})\!-\!f_{,R}^{(5)}R^{(5)}\right)g_{44}
-\left[\Box f_{,R}^{(5)} + \frac{\left(D_{\alpha}\Phi\right)\left(D^{\alpha}f_{,R}^{(5)}\right)}{\Phi} +\right.\right. \nonumber 
\\ 
&\left.\left.+  \frac{\epsilon}{\Phi^2}\left(\stackrel{\ast \ast}{f_{,R}^{(5)}} + \stackrel{\ast}{f_{,R}^{(5)}} \left(\frac{g^{\alpha\beta}\stackrel{\ast}g_{\alpha\beta}}{2} - \frac{\stackrel{\ast}\Phi}{\Phi}\right)\right) \right]g_{44}
+\epsilon\Phi\left(D_{\alpha}\Phi\right)\left(D^{\alpha}f_{,R}^{(5)}\right)+\stackrel{\ast\ast}{f_{,R}^{(5)}}-\frac{\stackrel{\ast}{\Phi}}{\Phi}\stackrel{\ast} {f_{,R}^{(5)}}\right\}\!.
\end{align}
\end{widetext}
Our purpose is now to disentangle the extra-coordinate dependence from the Einstein tensor. In this way it will be possible to isolate the 4-D part of Einstein tensor on the left member and move on the r.h.s all quantities that depend on higher derivative terms and on the extra-coordinate. All the higher order and higher dimensional counter-terms will play the role of effective source terms on the 4-D hypersurface. It is evident that since the fourth order Lagrangian $f(R^{(5)})$ depends on the 5-D Ricci scalar, this dependence cannot be completely untwined until the Lagrangian dependence is not specified. 
\\
The reduction rules for the Einstein tensor \cite{wessPdL92} give\: 

\begin{align}
\label{Riccimunu reduction}
R_{\mu\nu}^{(5)} = R_{\mu\nu} -& \frac{D_{\mu}D_{\nu}\Phi}{\Phi}+\frac{\epsilon}{2\Phi^2}\left(\frac{\stackrel{\ast}\Phi\stackrel{\ast}g_{\mu\nu}}{\Phi}-\stackrel{\ast\ast}g_{\mu\nu}+\right.\nonumber
\\
&\left.+g^{\lambda\alpha}\stackrel{\ast}g_{\mu\lambda} \stackrel{\ast} g_{\nu\alpha}\!-\!\frac{g^{\alpha\beta}\stackrel{\ast }g_{\alpha\beta}\stackrel{\ast}g_{\mu\nu}}{2}\right), 
\end{align}
\begin{align}\label{Ricci44 reduction}
R_{44}^{(5)} =\!-\!\epsilon \Phi \Box \Phi\!-\!\frac{\stackrel{\ast}g^{\alpha\beta}\stackrel{\ast}g_{\alpha\beta}}{4}\!-\!\frac{g^{\alpha\beta}\stackrel{\ast \ast}g_{\alpha\beta}}{2}\!+\!\frac{\stackrel{\ast}\Phi g^{\alpha\beta}\stackrel{\ast}g_{\alpha\beta}}{2\Phi}.
\end{align}
The next step is to calculate the 5-D Ricci scalar $R^{(5)}\!=\!g^{\mu\nu}R^{(5)}_{\mu\nu}+g^{44}R^{(5)}_{44}$. After the substitution of the expression (\ref{Ricci44 reduction}) within Eq.(\ref{curvquifRextra}) and some tedious algebra one gets the relation

\begin{align}\label{boxPhi}
\frac{\Box \Phi}{\Phi}=&\frac{\epsilon}{\Phi^2}\left(\frac{\stackrel{\ast}\Phi}{\Phi}\frac{g^{\alpha\beta}\stackrel{\ast}{g_{\alpha\beta}}}{2}-\frac{\stackrel{\ast}{g^{\alpha\beta}}\stackrel{\ast}{g_{\alpha\beta}}}{4}-\frac{g^{\alpha\beta}\stackrel{\ast\ast}{g_{\alpha\beta}}}{2}\right)+\nonumber
\\
-&\frac{1}{2}\frac{f(R^{(5)})}{f_{,R}^{(5)}}+\frac{\Box^{(5)}f_{,R}^{(5)}}{f_{,R}^{(5)}}-\frac{(D_{\alpha}\Phi)(D^{\alpha}f_{,R}^{(5)})}{\Phi f_{,R}^{(5)}}+
\\\nonumber
-&\frac{\epsilon}{\Phi^2}\left(\frac{\stackrel{\ast\ast}f_{,R}^{(5)}}{f_{,R}^{(5)}}-\frac{\stackrel{\ast}\Phi}{\Phi}\frac{\stackrel{\ast}{f_{,R}^{(5)}}}{f_{,R}^{(5)}}\right)\,,
\end{align}
that reminds an analogous expression given in \cite{deLeonjcap2010} for the 4-D reduction of 5-D Brans-Dicke theory. Further manipulations of (\ref{boxPhi}) allow to obtain a relatively simpler expression\,:

\begin{align}\label{boxPhi2}
\frac{\Box \Phi}{\Phi}=&\frac{\epsilon}{\Phi^2}\left(\frac{\stackrel{\ast}\Phi}{\Phi}\frac{g^{\alpha\beta}\stackrel{\ast}{g_{\alpha\beta}}}{2}-\frac{\stackrel{\ast}{g^{\alpha\beta}}\stackrel{\ast}{g_{\alpha\beta}}}{4}-\frac{g^{\alpha\beta}\stackrel{\ast\ast}{g_{\alpha\beta}}}{2}\right)+\nonumber \\
-&\frac{1}{2}\frac{f(R^{(5)})}{f_{,R}^{(5)}}+\frac{\Box f_{,R}^{(5)}}{f_{,R}^{(5)}}-\frac{\epsilon}{\Phi^2}\frac{g^{\alpha\beta}\stackrel{\ast}{g_{\alpha\beta}}}{2}\frac{\stackrel{\ast}{f_{,R}^{(5)}}}{f_{,R}^{(5)}}\,,
\end{align}
that can be, finally, used in the reduction of Ricci scalar to collect some terms. Therefore, by using also the trace equation, one has\;:

\begin{align}\label{curv5}
R^{(5)}&=R-2\frac{\Box\Phi}{\Phi}+\left(2\frac{\stackrel{\ast}\Phi}{\Phi}g^{\alpha\beta}\stackrel{\ast}{g_{\alpha\beta}}-
\right.\nonumber
\\
&\left. 2g^{\alpha\beta}\stackrel{\ast\ast}{g_{\alpha\beta}}-\frac{3}{2}\stackrel{\ast}{g^{\alpha\beta}}\stackrel{\ast}{g_{\alpha\beta}}-\frac{(g^{\alpha\beta}\stackrel{\ast}{g_{\alpha\beta}})^2}{2}\right).
\end{align}
It is easy to check that this result reproduces the analogous relation of 5-D GR \cite{KK} when $f(R^{(5)})\rightarrow R^{(5)}$. We remember that in such a case vacuum Einstein field equations reduce to the system $R_{AB}^{(5)}=0$ and $R^{(5)}$ vanishes.  

Once developed the several aspects of the 5-D to 4-D reduction procedure it is possible to obtain the induced 4-D field equations. By inserting (\ref{Riccimunu reduction}), (\ref{Ricci44 reduction}) and (\ref{curv5}) within Eqs. (\ref{curvquifRspazio})-(\ref{curvquifRextra}) 4-D spacetime equations get the form\footnote{The equations provided along this section partially reproduce previous results obtained in \cite{aguilar2015}. In fact, besides the different conceptual scheme, there are also some analytic differences that make the two schemes non completely indistinguishable.}:

\begin{equation}\label{space-time}
G_{\mu\nu}\,=\,T_{\mu\nu}^{IMT}+T_{\mu\nu}^{curv}+T_{\mu\nu}^{Mix}.
\end{equation}
Here: 

\begin{align*}
T_{\mu\nu}^{IMT}&=\frac{\Phi_{,\mu;\nu}}{\Phi}-\frac{\epsilon}{2\Phi^2}\left\{\frac{\stackrel{\ast}\Phi \stackrel{\ast}g_{\mu\nu}}{\Phi} - \stackrel{\ast \ast}g_{\mu\nu}+g^{\alpha\beta}\stackrel{\ast}g_{\mu\alpha}\stackrel{\ast}g_{\nu\beta}+\right.
\\
&\left.- \frac{g^{\alpha\beta}\stackrel{\ast}g_{\alpha\beta}\stackrel{\ast}g_{\mu\nu}}{2} + \frac{g_{\mu\nu}}{4}\left[\stackrel{\ast}g^{\alpha\beta}\stackrel{\ast}g_{\alpha\beta} + \left(g^{\alpha\beta}\stackrel{\ast}g_{\alpha\beta}\right)^2\right]\right\}.
\end{align*}
is the usual IMT tensor, deriving from the 5-D to 4-D reduction of Einstein tensor,

\begin{align}\label{curvquifR4}
 T^{Curv}_{\mu\nu}\!&=\!\frac{1}{f_{,R}^{(5)}}\left\{\frac{1}{2}\left(f(R^{(5)})-f_{,R}^{(5)}R\right)g_{\mu\nu}\!+\!
\right.\nonumber
\\
& \left.-\, [g_{\mu\nu}\,\Box-\nabla_\mu\nabla_\nu]\,f_{,R}^{(5)}\right\}\,.
\end{align}
is the curvature tensor written preserving the 4-D form. In such a case extra-dimension contributions are still hidden in the scalar curvature nested within the definition of the gravity Lagrangian and of its derivatives. Finally,

\begin{align}\label{Tmix}
T_{\mu\nu}^{Mix}\!\,=\,&\!\frac{\epsilon}{2\Phi^2}\stackrel{\ast}{g_{\mu\nu}}\frac{\stackrel{\ast}{f_{,R}^{(5)}}}{f_{,R}^{(5)}}-\frac{(D_{\alpha}\Phi)(D^\alpha f_{,R}^{(5)})}{\Phi f_{,R}^{(5)}}g_{\mu\nu}+\nonumber
\\
&-\frac{\epsilon}{\Phi^2}\left[\frac{\stackrel{\ast\ast}{f_{,R}^{(5)}}}{f_{,R}^{(5)}}+\frac{\stackrel{\ast}{f_{,R}^{(5)}}}{f_{,R}^{(5)}}\left(\frac{g^{\alpha\beta}\stackrel{\ast}{g_{\alpha\beta}}}{2}-\frac{\stackrel{\ast}\Phi}{\Phi}\right)\right]g_{\mu\nu}+
\\
&+\frac{\epsilon}{2\Phi^2}\left[\frac{\stackrel{\ast}{g^{\alpha\beta}}\stackrel{\ast}{g_{\alpha\beta}}}{4}+\frac{(g^{\alpha\beta}\stackrel{\ast}{g_{\alpha\beta}})^2}{4}\right]g_{\mu\nu}\nonumber
\end{align}
represents a mixed tensor containing terms that depend explicitly on the fifth coordinate and on the derivative of $f_{,R}^{(5)}$ with respect to $y$. 

%Actually, the three terms in the r.h.s. of Eqs. (\ref{space-time}) have only geometrical origin, descending respectively %from the extra-coordinate, the higher order derivative terms of the Lagrangian and their mixing.  in a top-down ``mode",

Therefore, we have obtained a framework where 4-D gravity is fully geometrized. In fact, within such a scheme matter-energy sources are related only with geometrical counter-effects deriving either from higher dimensional metric quantities or from higher order derivative ones. 

In order to complete our discussion one should take into account also the off-diagonal equation. By considering that $R^{(5)}_{4\alpha}\,=\,\Phi P_{\alpha\,;\beta}^{\beta}$ with $P_{\alpha\beta}=\displaystyle{\frac{1}{2\Phi}(\stackrel{\ast}{g_{\alpha\beta}}-g_{\sigma\rho}\stackrel{\ast}{g^{\sigma\rho}}g_{\alpha\beta})}$ \cite{wessPdL92}, one obtains 

\begin{equation}\label{0-4}
\Phi P_{\alpha\,;\beta}^{\beta} \!=\!\frac{\stackrel{\ast}{f_{,R\ ,\alpha}^{(5)}}}{f_{,R}^{(5)}}-\stackrel{\ast}{g_{\alpha\sigma}}g^{\beta\sigma}\frac{f_{,R\ ,\beta}^{(5)}}{f_{,R}^{(5)}}-\frac{\Phi_{,\alpha}}{\Phi}\frac{\stackrel{\ast}{f_{,R}^{(5)}}}{f_{,R}^{(5)}}
\end{equation}
that coincides with similar expressions achieved elsewhere \cite{deLeonjcap2010,aguilar2015} and in the limit $f(R^{(5)})\rightarrow R^{(5)}$ gives back, in absence of ordinary matter, the conservation law $P_{\alpha\ ;\beta}^\beta=0$ \cite{KK}. Equation (\ref{0-4}) resembles a more general conservation equation which relates the spacetime derivatives of $P_{\alpha}^{\beta}$ and $\stackrel{\ast}{f_{,R}^{(5)}}$.  It has been conjectured \cite{WessonP} that the spacetime components of the field equations relate geometry with the macroscopic properties of matter, while the extra-coordinate part ($^\alpha_4$)  and $(^4_4)$ might describe their microscopic ones. With these hypotheses, within fourth order gravity Kaluza Klein models, one would have that microscopic properties of matter are influenced by spacetime derivatives of the scalar degree of freedom intrinsic in $f(R)$ gravity. Furthermore, in our case, $P_{\alpha}^{\beta}$ dynamics depends also on $\Phi$, that, in turn (see Eq.(\ref{boxPhi2})), shows an evolution driven by the higher order gravity counter-terms. It seems that there is a strict interconnection between the extra-dimension properties and the intrinsic $f(R)$ scalar degree of freedom. For example, looking to equation (\ref{boxPhi2}), it seems that $f'(R)$ guarantees a sort of Machian effect for this kind of models. However, all these considerations represent, at this stage, nothing more than speculations and, therefore, we do not discuss this issue further here.

%%%%%%%%%%%%%%%%%%%%%%%%%%%%%%%%%%%%%%%%%%%%%%%%%%%
\section{A solving algorithm for 5-D f(R) gravity}
\label{geomat}
%%%%%%%%%%%%%%%%%%%%%%%%%%%%%%%%%%%%%%%%%%%%%%%%%%%%

\subsection{5-D f(R) gravity cosmological solutions}

Let us now verify what kind of solutions can be derived for our 5-D $f(R)$-gravity model. Once general cosmological solutions have been achieved, it is be possible to draw 5-D effects on $\Sigma_{y=y_0}$ hypersurfaces, deriving the effective 4-D picture. 
In particular, one can determine the spacetime effectual matter-energy behaviour of the different geometrical components described in the previous section. In order to search for field equations solutions we assume the 5-D metric:

\begin{align}\label{5Dmetric}
dS^2 =& n^2(t, y)dt^2 - a^2(t, y)\left[\frac{dr^2}{1 - k r^2}  + \nonumber \right. 
\\
&\left.+ r^2 \left(d\theta^2 + \sin^2 \theta d\varphi^2\right)\right] + \epsilon \Phi^2(t, y)dy^2,
\end{align}
where $k = 0, + 1, - 1$ is referred to the 3-D spacetime curvature and $(t, r, \theta, \phi)$ are the usual coordinates for spherically symmetric spatial sections. It is important to notice that, since the metric choice, our $f(R^{(5)})$ Lagrangian and its derivatives  do not depend on the 4-D spatial coordinates. In fact, the 5-D Ricci scalar calculated over the metric (\ref{5Dmetric}) is a function of the $(t, y)$ coordinates alone. If metric (\ref{5Dmetric}) is introduced in Eqs.(\ref{curvquifRspazio})-(\ref{curvquifRextra}), we obtain the equations system, respectively for the components: $_0^0$, $_i^i$ ($i\,=\,1,2,3$), $_4^4$ and $_4^0$\,:

\begin{widetext}
\begin{align}\label{5d-Eq00}
-\frac{f(R^{(5)})}{2f_{,R}^{(5)}}+\frac{3 \epsilon\stackrel{\ast}{a} \stackrel{\ast}{f_{,R}^{(5)}}}{\Phi^2 a f_{,R}^{(5)}}-\frac{3\epsilon \stackrel{\ast}{a} \stackrel{\ast}{n}}{\Phi^2 a n}-\frac{\epsilon\stackrel{\ast}{f_{,R}^{(5)}}\stackrel{\ast}{\Phi}}{f_{,R}^{(5)} \Phi^3}+\frac{\epsilon\stackrel{\ast}{n} \stackrel{\ast}{\Phi}}{n \Phi^3}+&\frac{\epsilon\stackrel{\ast\ast}{f_{,R}^{(5)}}}{\Phi^2 f_{,R}^{(5)}}-\frac{\epsilon\stackrel{\ast\ast}{n}}{\Phi^2 n} +\nonumber
\\
&+\frac{3 \dot{a} \dot{f_{,R}^{(5)}}}{a f_{,R}^{(5)} n^2}+\frac{3 \dot{a} \dot{n}}{a n^3}+\frac{\dot{f_{,R}^{(5)}} \dot{\Phi}}{n^2 f_{,R}^{(5)} \Phi}+\frac{\dot{n} \dot{\Phi}}{n^3 \Phi}-\frac{3 \ddot{a}}{n^2a}-\frac{\ddot{\Phi}}{n^2 \Phi}=0\,,
\end{align}

\begin{align}\label{5d-Eqii}
-\frac{f(R^{(5)})}{2 f_{,R}^{(5)}}-\frac{2 \epsilon\stackrel{\ast}{a}^2}{\Phi^2 a^2}+\frac{2 \epsilon \stackrel{\ast}{a} \stackrel{\ast}{f_{,R}^{(5)}}}{\Phi^2 a f_{,R}^{(5)}}-\frac{\epsilon\stackrel{\ast}{a} \stackrel{\ast}{n}}{\Phi^2 a n  }&+\frac{\epsilon\stackrel{\ast}{f_{,R}^{(5)}} \stackrel{\ast}{n}}{\Phi^2 f_{,R}^{(5)}n}+\frac{\epsilon\stackrel{\ast}{a} \stackrel{\ast}{\Phi}}{a \Phi^3}-\frac{\epsilon f_{,R}^{(5)} \stackrel{\ast}{\Phi}}{\Phi^3 f_{,R}^{(5)}}-\frac{\epsilon\stackrel{\ast\ast}{a}}{\Phi^2 a }+\frac{\epsilon\stackrel{\ast\ast}{f_{,R}^{(5)}}}{\Phi^2 f_{,R}^{(5)}}+\nonumber
\\
&-\frac{2 \dot{a}^2}{a^2 n^2}+\frac{2 \dot{a} \dot{f_{,R}^{(5)}}}{a f_{,R}^{(5)} n^2}+\frac{\dot{a} \dot{n}}{a n^3}-\frac{\dot{f_{,R}^{(5)}} \dot{n}}{f_{,R}^{(5)} n^3}-\frac{\dot{a} \dot{\Phi}}{a n^2 \Phi }+\frac{\dot{f_{,R}^{(5)}} \dot{\Phi}}{f_{,R}^{(5)} n^2 \Phi }-\frac{\ddot{a}}{a n^2}+\frac{\ddot{f_{,R}^{(5)}}}{f_{,R}^{(5)} n^2}=0\,,
\end{align}

\begin{align}\label{5d-Eq55}
-\frac{f(R^{(5)})}{2 f_{,R}^{(5)}}+\frac{3 \epsilon \stackrel{\ast}{a} \stackrel{\ast}{f_{,R}^{(5)}}}{\Phi^2 a f_{,R}^{(5)}}+\frac{\epsilon\stackrel{\ast}{f_{,R}^{(5)}} \stackrel{\ast}{n}}{\Phi ^2 f_{,R}^{(5)} n}+\frac{3\epsilon \stackrel{\ast}{a} \stackrel{\ast}{\Phi}}{a \Phi^3}+\frac{\epsilon\stackrel{\ast}{n} \stackrel{\ast}{\Phi}}{n \Phi^3}&-\frac{3\epsilon \stackrel{\ast\ast}{a}}{\Phi^2 a }-\frac{\epsilon\stackrel{\ast}{n}}{\Phi^2 n}+\nonumber
\\
&+\frac{3 \dot{a} \dot{f_{,R}^{(5)}}}{a f_{,R}^{(5)} n^2}-\frac{\dot{f_{,R}^{(5)}} \dot{n}}{f_{,R}^{(5)} n^3}-\frac{3 \dot{a}\dot{\Phi}}{a n^2 \Phi}+\frac{\dot{n} \dot{\Phi} }{n^3 \Phi}+\frac{\ddot{f_{,R}^{(5)}}}{f_{,R}^{(5)} n^2}-\frac{\ddot{\Phi}}{n^2 \Phi}=0\,,
\end{align}

\begin{equation}\label{5d-Eq04}
\frac{3 \dot{a} \stackrel{\ast}{n}}{a n}+\frac{ \stackrel{\ast}{n} \dot{f_{,R}^{(5)}}}{n f_{,R}^{(5)} }+\frac{3 \stackrel{\ast}{a} \dot{\Phi}}{ a  \Phi }+\frac{ \stackrel{\ast}{f_{,R}^{(5)}} \dot{\Phi}}{f_{,R}^{(5)} \Phi}-\frac{3 \dot{\stackrel{\ast}{a}}}{ a }-\frac{ \dot{\stackrel{\ast}{f_{,R}^{(5)}}}}{f_{,R}^{(5)}}=0\,.
\end{equation}

\end{widetext}
The last equation (\ref{5d-Eq04}) recalls, again, the result obtained in \cite{deLeonjcap2010} for 5-D Brans-Dicke vacuum solutions. Therefore it seems a suitable choice to pursue the same approach performed in this work to get some particular solutions of our equations system (\ref{5d-Eq00})-(\ref{5d-Eq04}). If we consider the metric coefficients as separable functions of their arguments\,:

\begin{align}\label{metric separation}
n(t, y) = N(y),\;\;\; a(t, y) = P(y)Q(t),\;\;\;\nonumber
\\
\Phi(t, y) = F(t),\;\;\; f_{,R}^{(5)}(t, y) = U(y)W(t),
\end{align}
from Eq.$^0_4$ one gets\,:

\begin{equation}
\frac{3 \dot{F} \stackrel{\ast}{P}}{F P}+\frac{3 \stackrel{\ast}{N} \dot{Q}}{N Q}-\frac{3 \stackrel{\ast}{P} \dot{Q}}{ P Q}+\frac{ \dot{F} \stackrel{\ast}{U}}{F U}+\frac{\stackrel{\ast}{N} \dot{W}}{N W}-\frac{\stackrel{\ast}{U} \dot{W}}{U W}=0\,.
\end{equation}
This equation suggests the possibility to look for power law cosmological solutions. In the following section we will discuss some examples  in this sense.
%, we leave to a forthcoming paper a more detailed analysis about cosmological power law solutions in the framework of 5-D $f(R)$ gravity.

\subsection{Preliminary power law solutions}

In order to look for 5-D $f(R)$ gravity power law solutions we consider a set of metric potentials defined as follows \cite{deLeonjcap2010}:

\begin{equation}
\label{plawpotenziali}
n(t, y) = N_0 y^{\delta}, \;\;\;a(t, y) = A_0 t^{\alpha} y^{\beta}, \;\;\;\Phi(t, y) = \Phi_0 t^{\gamma}y^{\sigma}\,, 
\end{equation}
with $N_0, A_0, \Phi_0$ some constants with appropriate units and  $\alpha, \beta, \gamma, \delta, \sigma$  that represent the unknown parameters asked to satisfy 5-D field equations. 
\\
Up to now we have developed a completely general scheme, without any assumption on the gravity Lagrangian. However, to completely define the solving algorithm one has to make a choice about the Ricci scalar function entering the gravity action. To remain conservative with the solution procedure, we consider a power law function of the Ricci scalar  $f(R)=f_0 R^n$. Such a model has been extensively studied in literature at four dimensions both in cosmology (curvature quintessence) \cite{curv1,curv2,curv3} and in the low energy limit \cite{curvlow,curvgalaxies}. In addition, the 5-D phenomenology of a power law fourth order gravity has been investigated in the standard approach in presence of a perfect fluid matter source \cite{huang2010,wu2014}. In particular, in such a case, it is assumed a compact fifth dimension and it is supposed that the homogeneously distributed fluid does not travel along the fifth dimension. As a matter of fact the matter energy density and the pressure experienced by a four-dimensional observer have to be an integrated throughout the compact extra-dimensional ring. In our approach we have overcome this point of view, we relax the hypotheses on the extra-coordinate and, above all, we discard ordinary matter in favor of a completely geometric self-consistent approach. According with cosmological observations  \cite{cmb,planck}, we will study field equations considering a flat spacetime geometry ($k=0$).  

A suitable recipe to study Eqs. (\ref{5d-Eq00})-(\ref{5d-Eq04}), is to plug the metric functions (\ref{plawpotenziali}) within the last of these relations and to search for its solutions. This approach allows to obtain, with some efforts, a number of constraints on the model parameters. Then, one can try to satisfy also the other more complicated field equations (\ref{5d-Eq00})-(\ref{5d-Eq55}). By inserting metric potentials within Eq.$^0_4$, one obtains a rather cumbersome expression\,:

\begin{equation}\label{04_pl}
\frac{\Xi(\alpha,\beta,\gamma,\delta,\sigma,n,t,y)}{\Upsilon(\alpha,\beta,\gamma,\delta,\sigma,n,t,y)}\,=\,0
\end{equation}
where

\begin{widetext}
\begin{multline*}
\Xi(\alpha,\beta,\gamma,\delta,\sigma,n,t,y)\,=\,y^{-1-2 \delta} \Big(3 (\beta  \gamma +\alpha  (-\beta +\delta )) \left(N_0^2 t^2 y^{2 \delta } \left(6 \beta ^2+3 \beta  (-1+\delta -\sigma )+\delta  (-1+\delta -\sigma )\right)+\right.\\
\left.\left.+ t^{2 \gamma } y^{2+2 \sigma } \left(6 \alpha ^2+3 \alpha  (-1+\gamma )+(-1+\gamma ) \gamma \right) \epsilon  \Phi_0^2\right)^2-\right.\\
\left.+2 (-1+n) \left(N_0^4 t^4 y^{4 \delta } \gamma  (\delta +(-1+2 n ) (1+\sigma )) \left(-6 \beta ^2+3 \beta  (1-\delta +\sigma )+\delta  (1-\delta +\sigma )\right)^2+\right.\right.\\
\left.\left.+ N_0^2 t^{2+2 \gamma } y^{2 (1+\delta +\sigma )} \left(6 \alpha ^2+3 \alpha  (-1+\gamma )+(-1+\gamma ) \gamma \right) \epsilon  \left(6 \beta ^2+3 \beta  (-1+\delta -\sigma )+\right.\right.\right.\\
\left.\left.\left. +\delta  (-1+\delta -\sigma )\right) (3 \delta +2 (-2+n ) (1+\sigma )+\gamma  (3-2 \delta +2 \delta  n +3 \sigma )) \Phi_0^2+\right.\right.\\
\left.+ t^{4 \gamma } y^{4+4 \sigma } \left(6 \alpha ^2 +3 \alpha  (-1+\gamma )+(-1+\gamma ) \gamma \right)^2 \delta  \epsilon ^2 (-1+\gamma +2 n) \Phi_0^4\right) \Big)
\end{multline*}
and

\begin{multline*}
\Upsilon(\alpha,\beta,\gamma,\delta,\sigma,n,t,y)\,=\,N_0^2 t \left( N_0^2 t^2 y^{2 \delta } \left(6 \beta ^2+3 \beta  (-1+\delta -\sigma )+\delta  (-1+\delta -\sigma )\right) \right.\\
\left. + t^{2 \gamma } y^{2+2 \sigma} \left(6 \alpha ^2+3 \alpha  (-1+\gamma )+(-1+\gamma ) \gamma \right) \epsilon  \Phi_0^2\right)^2.
\end{multline*}
\end{widetext}
This result, evidently, suggests the possibility to obtain, in principle, much more solutions than the Brans-Dicke case studied in \cite{deLeonjcap2010}. To simplify our search, as a preliminary step, we make some trivial hypotheses about model parameters, leaving a complete analysis of field equations solutions to a future dedicated work. 

\subsubsection{Kaluza-Klein GR limit}

As a first step we want to verify that the Kaluza-Klein GR limit is recovered. To perform standard KK limit we have to settle $n=1$ and, in addition, following customary approaches to the model, we assume $\sigma=0$.

In such a case Eq.(\ref{04_pl}) becomes very simple and can be verified if $\beta  \gamma +\alpha  (-\beta +\delta )=0$. In particular, if $\gamma\neq\alpha$ one obtains $\beta \to \displaystyle\frac{\alpha  \delta }{\alpha -\gamma }$ (a result that is in agreement with \cite{deLeonjcap2010}). Looking to the other equations, together some trivial non evolving solutions ($\alpha=0$\footnote{It is possible to find some trivial solutions with, $\alpha=0$, $\beta=0$, $\delta=(0,1)$, $\gamma=1$}), one obtains the solving sets of parameters\,:

\begin{align}\label{GRsol1}
\delta = 1,\,\gamma=1,\,\alpha= &-1,\,\beta=1 \nonumber\\
 \text{and}\  N_0^2=4\Phi_0^2\  &\text{with}\  \epsilon =-1\,,
\end{align} 

\begin{align}\label{GRsol2}
\delta = 0,\,\gamma=-1/2,\,\alpha= &1/2,\,\beta=0\,,
\end{align} 
with $\epsilon=\pm1$ (no restrictions on the y coordinate), and

\begin{align}\label{GRsol3}
\delta = 1,\,\gamma=1,\,\alpha= &-1,\,\beta=\frac{\alpha}{\alpha-1} \nonumber\\
 \text{and}\  \Phi_0^2=\frac{N_0^2}{(\alpha-1)^2}\  &\text{with}\  \epsilon =-1\,.
\end{align} 
All these solutions are well known in literature \cite{deLeonjcap2010,PdL88}, therefore, our model completely reproduces standard GR Kaluza-Klein models in the $f(R^{(5)}) \rightarrow R^{(5)}$ limit.

\subsubsection{Standard 4-D $f(R)$ gravity limit}

What about standard 4-D fourth order gravity? One should expect that this framework has to represent a natural subcase of 5-D $f(R)$ gravity. In order to get this limit one has to assign a solutions set with $\delta=0,\,\beta=0,\,\sigma=0$ and $\gamma=0$, that means, no dependence on $y$ and  no dynamics on the fifth coordinate. Starting with this assumptions and by considering equations (\ref{5d-Eq00})-(\ref{5d-Eqii}) and (\ref{5d-Eq04}),  one obtains

\begin{equation}\label{fR4-D}
 \alpha=\frac{-1+3n-2n^2}{-2+n},
\end{equation} 
that exactly matches the well known power law $f(R)$ gravity solution in the case of vacuum field equations \cite{curv1}. However, in the 5-D case, to fully satisfy the theory we have to fulfill an equation more, Eq. (\ref{5d-Eq55}). This request,  by fact, settles the power law index $n$. Therefore the only allowed combination is $\delta=0,\,\beta=0,\,\sigma=0,\,\gamma=0,\,n=5/4$ and $\alpha=1/2$. 
 
\subsubsection{Cylinder-type solutions}

 We can now relax some hypotheses in order to look for generalizations of 5-D GR. To search for such a kind of solutions we start, as a first simple case, from the assumption of no dependence on the extra-coordinate. This means to study our theory, looking for solutions of Eqs.(\ref{5d-Eq00})-(\ref{5d-Eq04}), in the limit of cylinder condition. Differently than the standard case, in the $f(R)$ generalization of Induced Matter Theory this condition, as we will see, does not imply radiation as the only possible kind of induced matter. 
 
At first, let us gibe a look to Eq.(\ref{boxPhi2}). In the case of 5-D $f(R)$ gravity, independence on the fifth coordinate does not mean to have a massless Klein-Gordon equation for the extra-coordinate potential $\Phi$. Higher order gravity counter-terms will play the role of a mass term and $\Phi$ will play a completely new role into the dynamics\,:
 
 \begin{align}\label{boxPhi_KG}
\Box \Phi=-\left[\frac{1}{2}\frac{f(R^{(5)})}{f_{,R}^{(5)}}+\frac{\Box f_{,R}^{(5)}}{f_{,R}^{(5)}}\right]\Phi .
\end{align}
In addition, as a consequence, $T_{\mu\nu}^{IMT}$ will be no more traced to zero \cite{KK}, therefore the Induced Matter Tensor can span more general kinds of matter in relation to the underlying cosmological solution. 

From the  field equations point of view, power law solutions of cylinder-type are achieved when the model parameters in (\ref{plawpotenziali}) are settled as: $\delta=0,\,\beta=0,\,\sigma=0$ with free $n$. Neglecting again static solutions\footnote{There are also in this case trivial solutions i.e. $n=1$, $\alpha=0$, $\gamma=(0,\,1)$.} it is possible to find a set of implicit solutions that can be expressed in term of the $f(R)$ Lagrangian power index $n$\,:

\begin{widetext}
\begin{subequations}
\begin{align}
\alpha_1 = \frac{1}{12} \left(-3+6 n -f(n)\right),\ &\gamma_1 = -\frac{17+28 n ^2+3 f(n)-2 n  \left(24+f(n)\right)}{2 \left(-5+2 n +f(n)\right)}\,;\label{solcyl1}\\
\alpha_2 = \frac{1}{12} \left(-3+6 n +f(n)\right),\ &\gamma_2 = \frac{17+28 n ^2-3 f(n)+2 n  \left(-24+f(n)\right)}{2 \left(5-2 n +f(n)\right)}\,;\label{solcyl2}\\
\alpha_3 = &\gamma_3 = \frac{2-6 n +4 n ^2}{5-2 n }\label{solcyl3}\,;
\end{align}
\end{subequations}
\end{widetext}
\noindent with $f(n)=\sqrt{-39+108 n -60 n ^2}$.  Eqs.(\ref{solcyl1})-(\ref{solcyl3}) represent an interesting achievement. In the following we will show that these solutions provide, after the 5-D to 4-D reduction, cosmological significant behaviours of the 4-D matter-energy tensors induced by 5-D geometry.

%%%%%%%%%%%%%%%%%%%%%%%%%%%%%%%%%%
\section{4-D induced $f(R)$ gravity: a top-down geometrization of matter}
%%%%%%%%%%%%%%%%%%%%%%%%%%%%%%%%%%

The effective 4-D picture induced by higher dimensions can be achieved once cosmological solutions obtained in Sec.\ref{geomat} are plugged within the 5-D to 4-D reduction framework previously outlined. In particular the 4-D setting is achieved, as already said, by considering hypersurfaces $\Sigma_{y=y_0}$ that slice the 5-D universe along the fifth coordinate. In this scheme, the extra-coordinate effects on 4-D physical quantities are evaluated taking $y=y_0=const$. The resulting cosmological model is a 4-D induced matter $f(R)$ theory, where field equations are given in the form of Einstein equations equipped with three different matter-energy sources of geometrical origin. Deviations from standard GR provide, in the effective 4-D spacetime description, matter-energy sources and one could wonder if these contributions are related with the elusive nature of Dark Energy and Dark Matter. Actually, we study the 4-D induced cosmologies in correspondence of the solutions given in Sec.\ref{geomat}. In particular, we are interested to investigate the behaviour of the three different induced cosmological fluids  (\ref{IMT}), (\ref{curvquifR4}), (\ref{Tmix}) on the $y=y_0=const$ hypersurfaces. One can notice that for all of these quantities it holds $T_1^{1\,K}=T_2^{2\,K}=T_3^{3\,K}$ (with $K$=$IMT$, $Curv$, $Mix$), therefore it is possible to describe such matter-energy tensors as perfect fluid sources with $\rho_K=T_0^{0\ K}$ and the isotropic pressure defined as $p_K=-T_1^{1\,K}$. The cosmological nature of each component can be evaluated by means of the respective equation of state (EoS) $\omega_K=-T_1^{1\,K}/T_0^{0\,K}$.

\subsection{4-D GR limit}

If one considers solutions obtained along the standard KK GR limit, the related 4-D effective dynamics are described by well known cosmological models. In particular, the solution (\ref{GRsol2}) represents the standard radiation universe with $p=1/3\rho$. In such a case the three matter-energy tensors collapse into one matter-energy source, in particular the curvature quintessence tensor and the mixed tensor respectively vanish. On the other side, the solution (\ref{GRsol3}) represents a spatially flat FRW metric once the extra-coordinate is fixed\,:

\begin{align} 
ds^2_{|\Sigma_y}=N_0^2y_0^2dt^2-A_0^2y_0^{\frac{2\alpha}{\alpha-1}}t^{2\alpha}\left[dr^2+r^2(d\theta^2+\sin^2d\varphi)\right].
\end{align}
The induced matter tensor plays, again, the role of the only one effective cosmological fluid. In fact, the other two quantities combine to give a  cancelling result. The cosmological energy density scales as $\rho=\rho_0t^{-2}$, with $\rho_0=\displaystyle\frac{3 \alpha ^2}{N_0^2 y_0^2}$, while the barotropic factor is $\omega=\displaystyle\frac{2-3\alpha}{3\alpha}$. This result is in complete agreement with previous achievements on non compactified KK gravity \cite{deLeonjcap2010}.

\subsection{4-D cosmology from standard $f(R)$ gravity limit}

We have seen that the attempt to get the standard 4-D $f(R)$ gravity limit is frustrated by the higher equations number of our  model. In fact, it is possible to get the right 4-D power law solution only if the extra-coordinate equation is neglected. When this equations enters the game the power law index $n$ gets fixed. The behaviour of the three matter-energy tensors describing the top-down effect of higher dimensions and higher derivative terms on the 4-D gravity hypersurface confirmate such a result. In fact, if we evaluate $\omega_{curv}$ on the 4-D $f(R)$ solution (\ref{fR4-D}) we obtain 

\begin{equation}
\omega_{curv}=\frac{1+7n+6n^2}{3-9n+6n^2},
\end{equation}
that, again, matches the already known result for this kind of model \cite{curv1}. However, since the fifth equation select the allowed values for $n$, we know that the admissible solution is indeed $\delta=0,\,\beta=0,\,\sigma=0,\,\gamma=0,\,n=5/4$ and $\alpha=1/2$. The effective, 4-D, consequent cosmology shows $\rho_{IMT}=\rho_{Mix}=0,\,\rho_{curv}=\frac{3}{4N_0^2 t^{2}}$ and $\omega_{curv}=1/3$.  As a matter of fact, we find a new radiation like universe with $n=5/4$. The radiation component derivates from the higher derivatives counter-terms while the Induced Matter Tensor vanishes.

\subsection{4-D Cylinder models}

\begin{figure*}[htbp]
\begin{center}
\includegraphics[width=12cm]{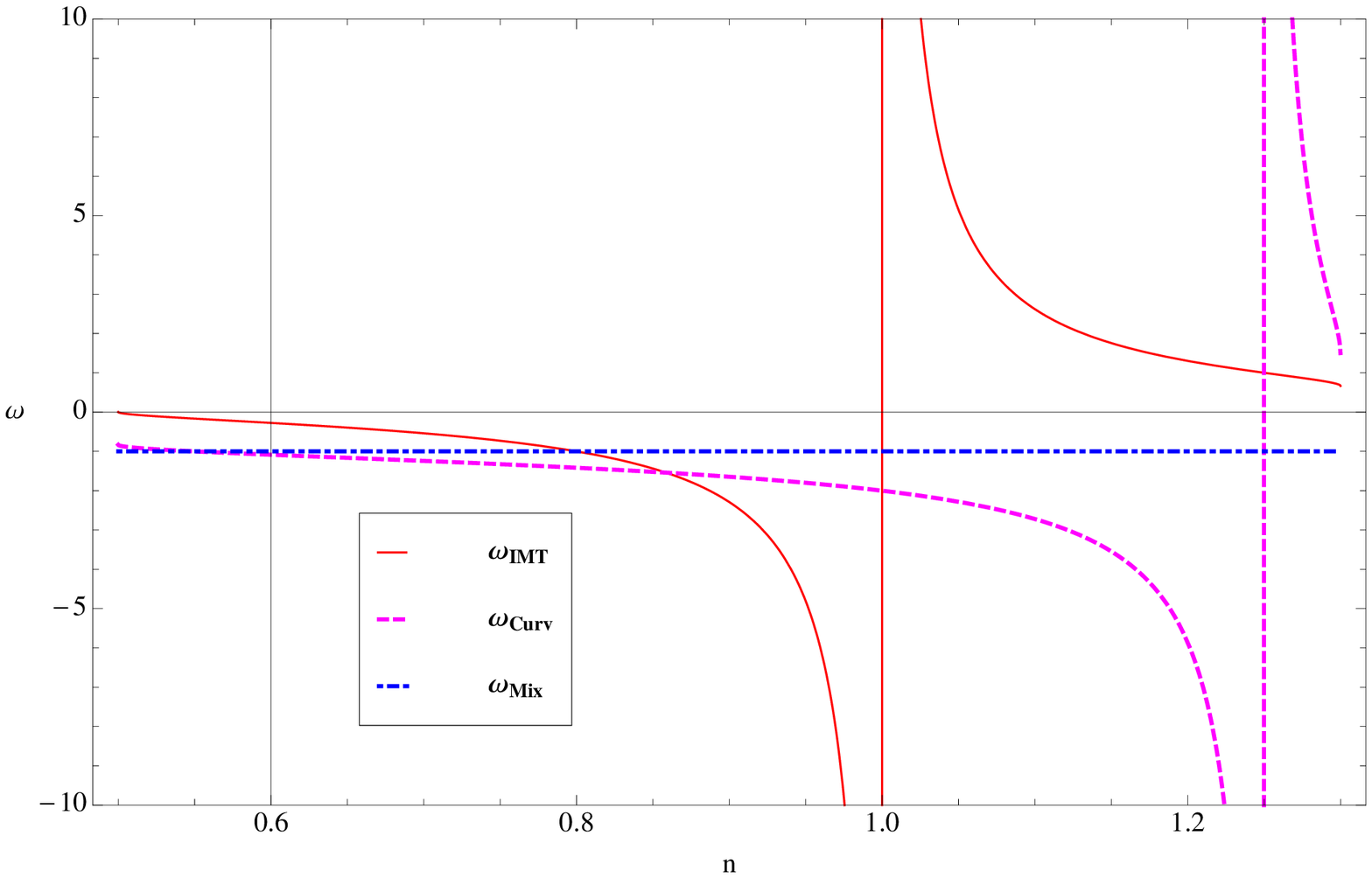}
\includegraphics[width=12cm]{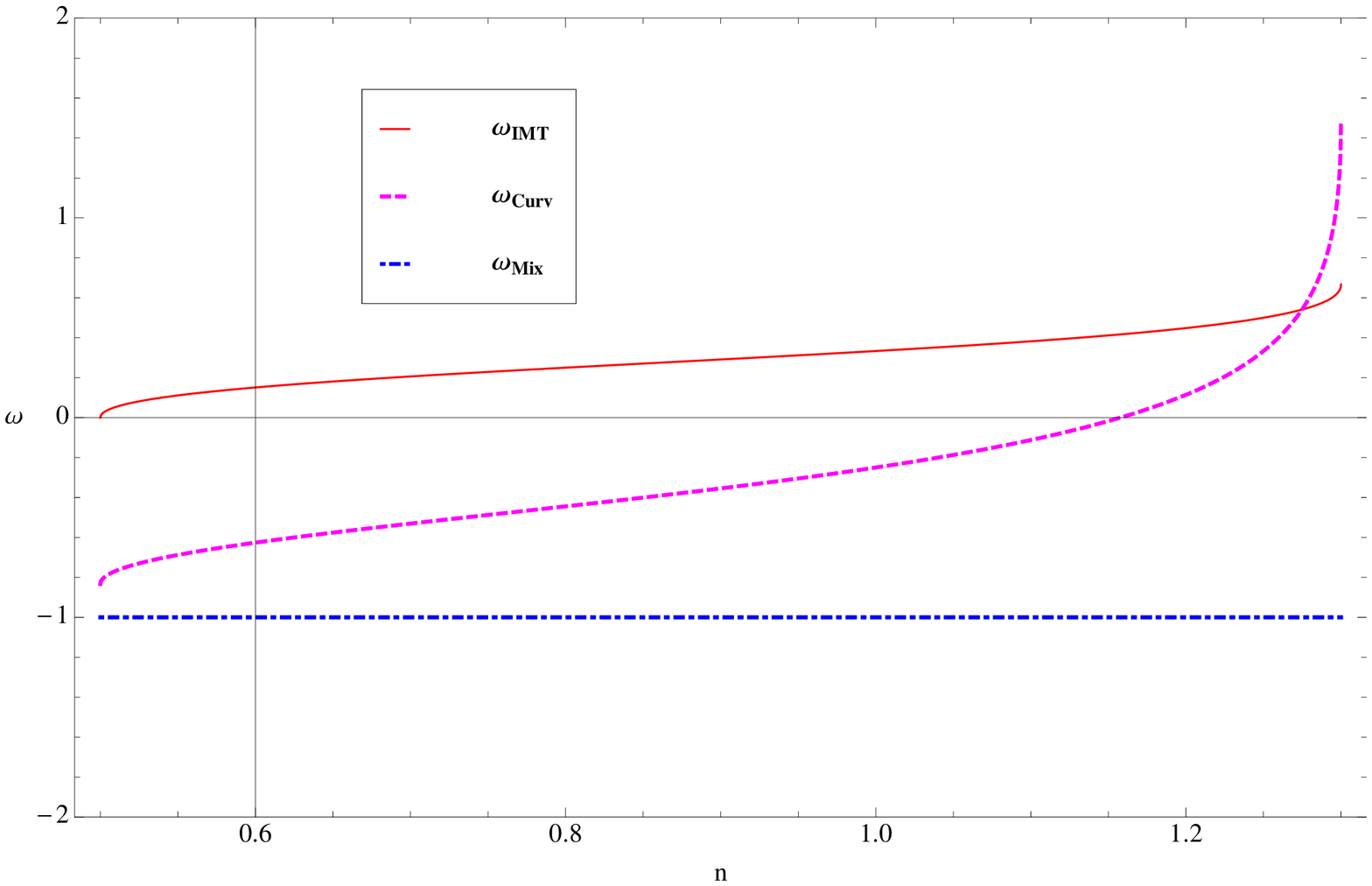}
\includegraphics[width=12cm]{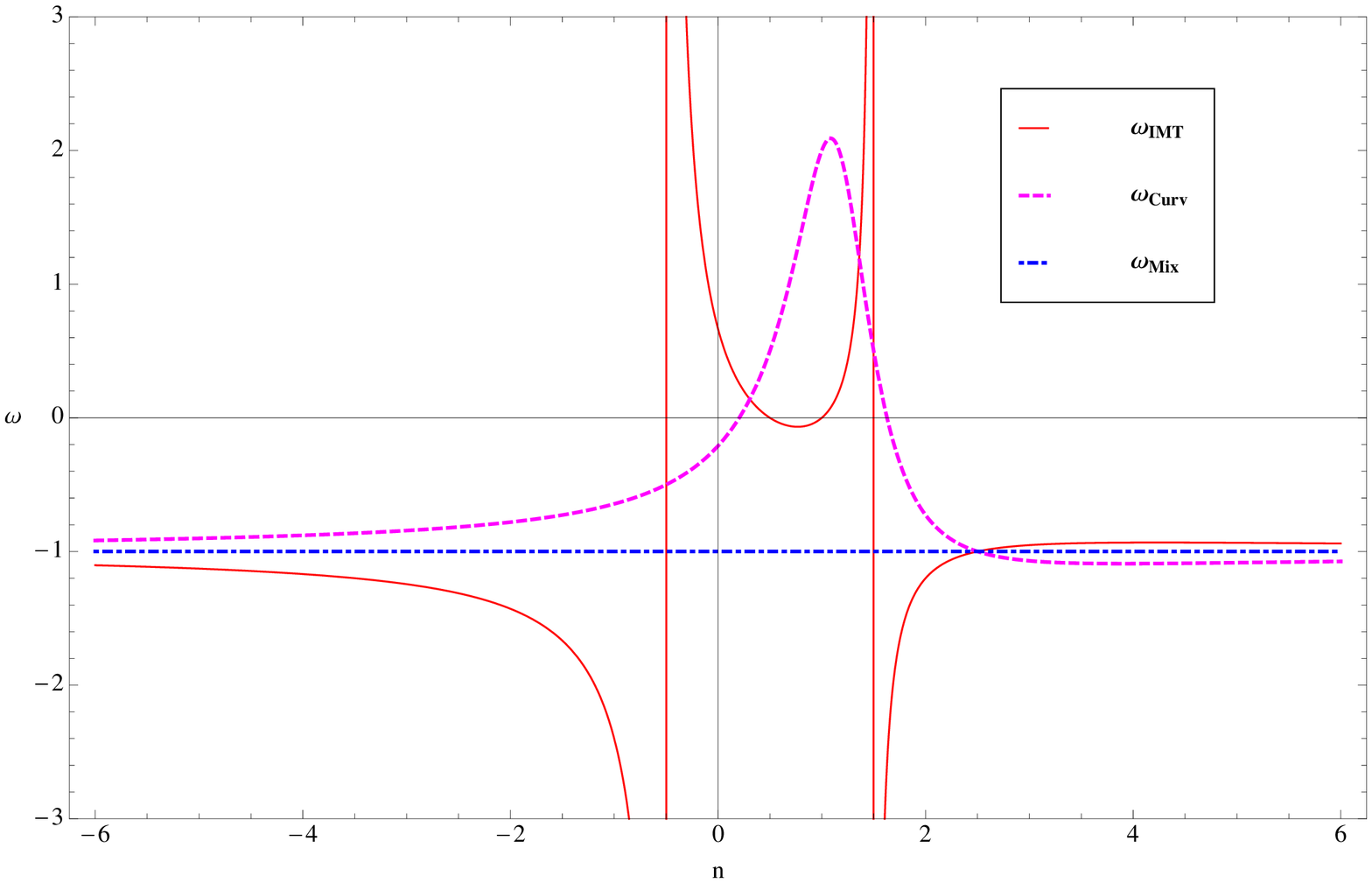}
\caption{The behaviour of the barotropic factors related to three matter-energy tensors  that arise after the 5-D to 4-D reduction. The three panel refer to the EoS comportment with respect to $n$ along each of the cylinder solutions (\ref{solcyl1})-(\ref{solcyl3}) (from top to down).}
\label{figure1}
\end{center}
\end{figure*}

Cosmologies induced by cylinder solutions are quite interesting and allow different dynamics. In particular, the matter-energy sector, characterized by means of the three matter-energy sources of geometrical origin, assumes a variety of significant behaviours. A relevant aspect of these solutions is that the mixed tensor $T^{Mix}_{\mu\nu}$ provides a cosmological constant-like source. In fact, as it is possible to observe from the definition (\ref{Tmix}), in such a case (no extra-coordinate dependence) it is $T^{0\,Mix}_0=T^{i\,Mix}_i$, therefore $\omega_{Mix}=-1$. Since for each solution, $\rho_{Mix}=T^{0\,Mix}_0\sim X_i(n){N_0t}^{-2}$, with $X_i(n)$ depending on the solution we consider, we obtain an inverse square law cosmological constant. This behaviour is favored in string cosmologies \cite{lopez1996} and in time-varying $\Lambda$ theories that accommodate a large $\Lambda$ for early times and a negligible $\Lambda$ for late times \cite{overcoop1998}. On the other side, the effective barotropic factors of the other two matter-energy sources depend on the parameter $n$ in relation to the solutions (\ref{solcyl1})-(\ref{solcyl3}). In order to have a schematic portrait of these result we have plotted the behaviour of each $\omega_{K}$ with respect to $n$ in Fig.\ref{figure1}. It is possible to observe that the value of the gravity Lagrangian power index $n$ determines the possibility to have different cosmological components.  Both the standard matter sector and the dark energy one can be mimicked by the geometrically induced fluids  according with previous results in this sense obtained in the framework of IMT \cite{wesson1992,liu9298,socorro1996} and fourth order gravity \cite{curvmimick}. Very interesting is the coexistence of different regimes with the same value of $n$.

\begin{figure*}[htbp]
\begin{center}
\includegraphics[width=12cm]{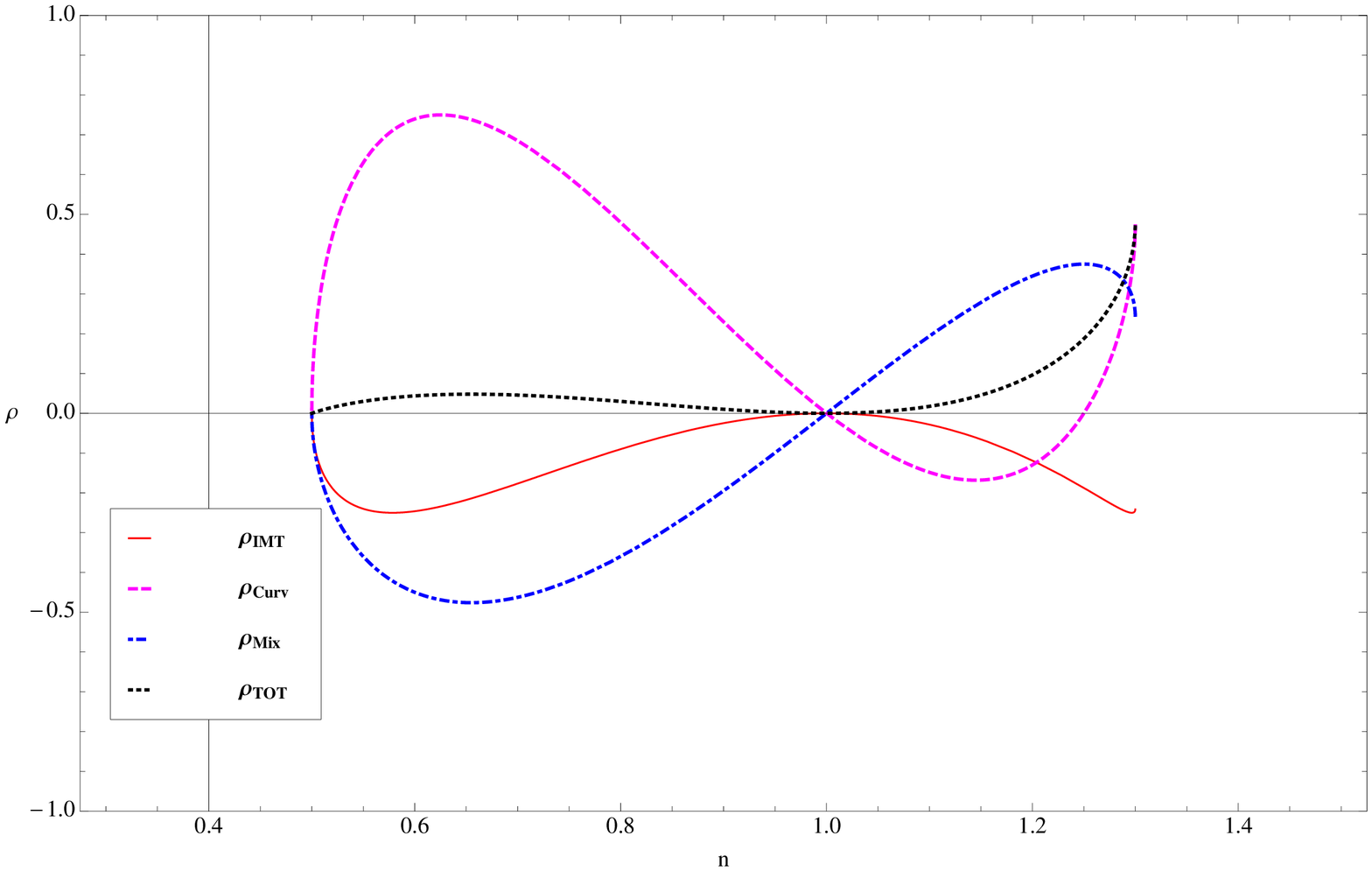}
\includegraphics[width=12cm]{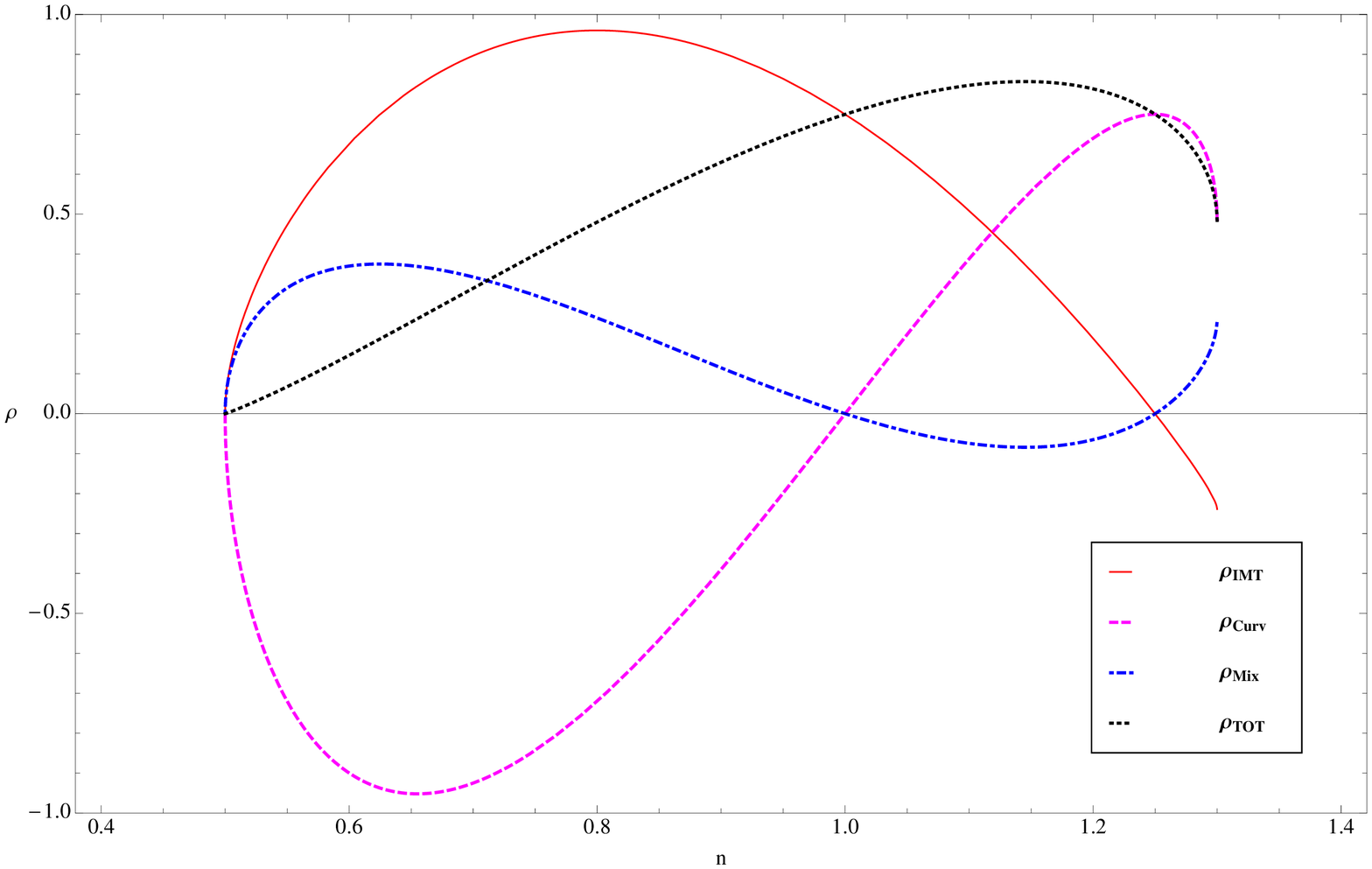}
\includegraphics[width=12cm]{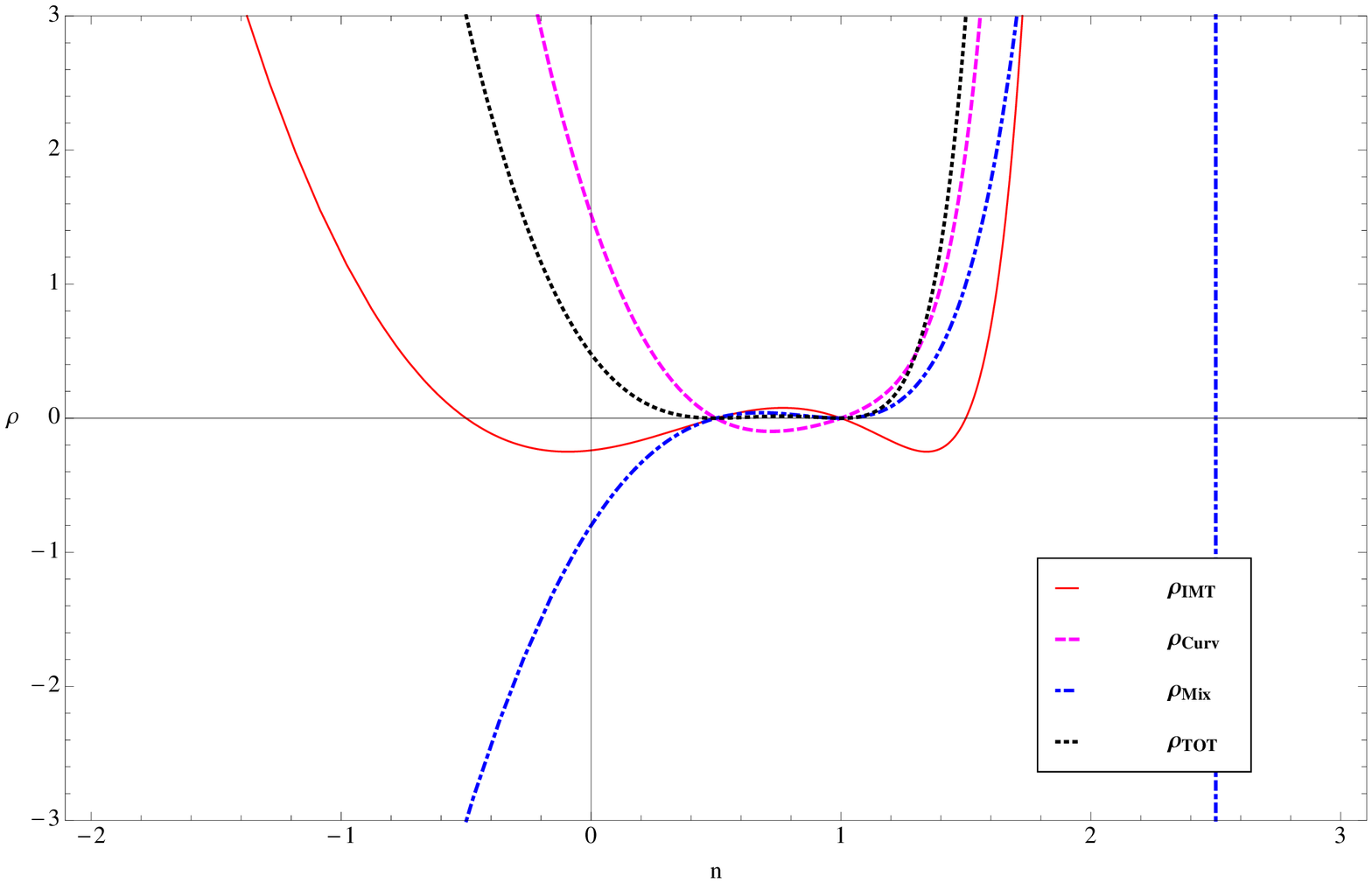}
\caption{Energy densities for the different species of matter-energy obtained in a 5D f(R)-gravity model when reduced to the standard 4-D spacetime. Curves labelled with $\rho_{tot}$ describe the behaviour of the whole matter-energy cosmological budget; natural units are adopted. The upper panel refers to solution (\ref{solcyl1}), the middle one to (\ref{solcyl2}), finally the last one to solution (\ref{solcyl3}).}
\label{figure2}
\end{center}
\end{figure*}

\begin{figure*}[htbp]
\begin{center}
\includegraphics[width=12cm]{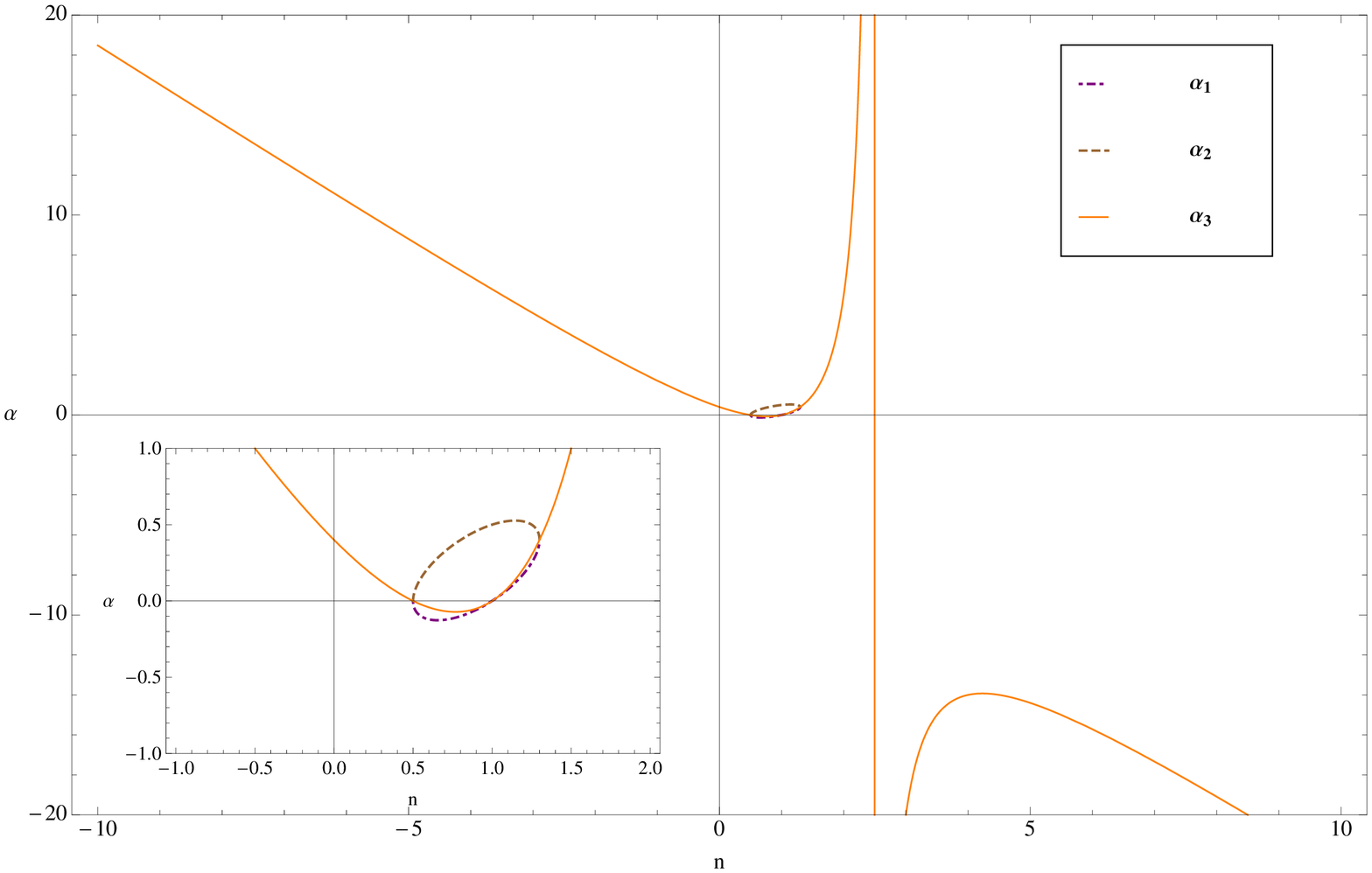}
\includegraphics[width=12cm]{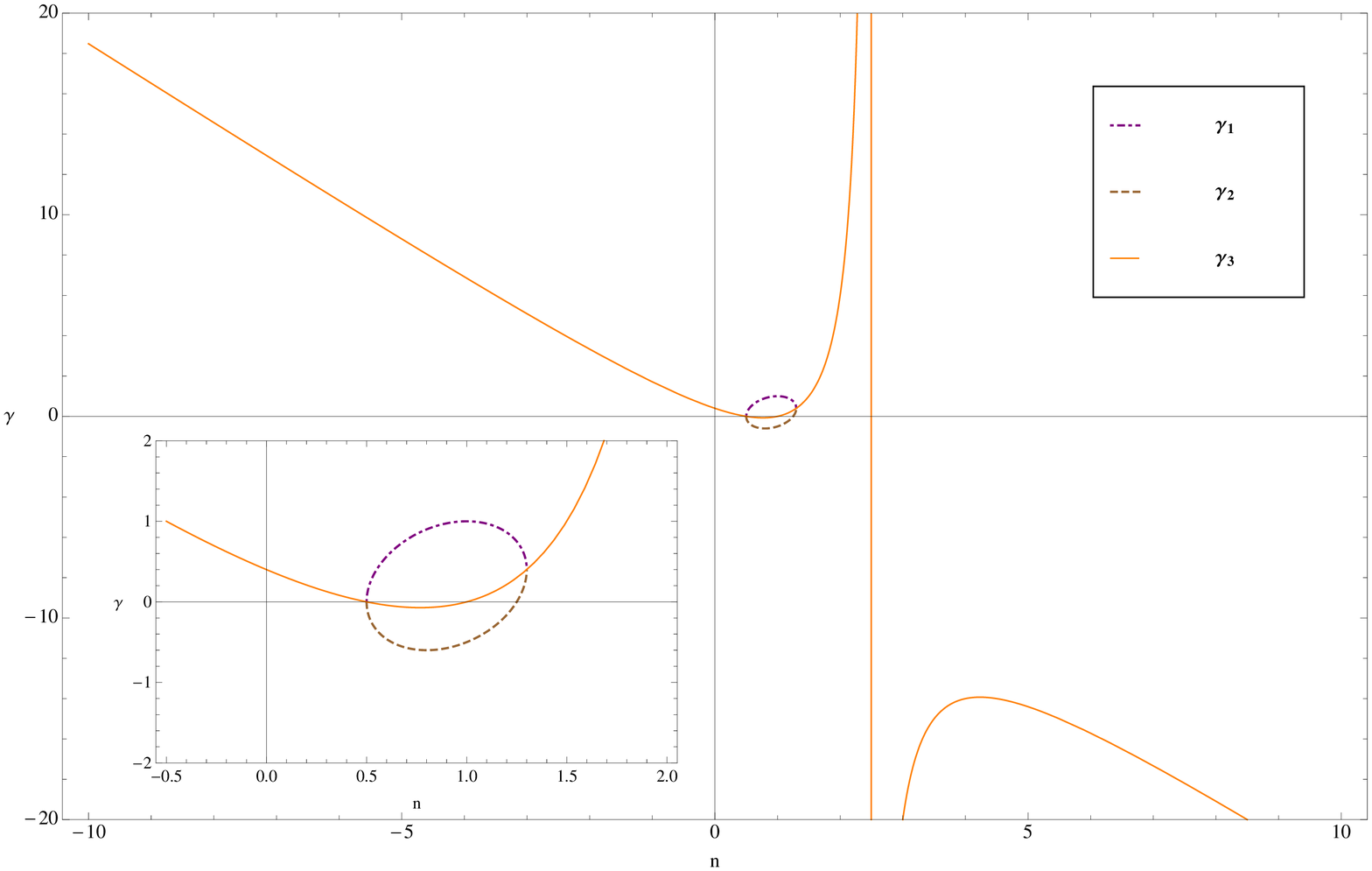}
\caption{The scale factor and the extra coordinate power law indexes vs. $n$ for the three cylinder solutions obtained along the 5-D to 4-D reduction of higher order gravity power law models. Both plots contain a detail that enlightens the behaviour in the region $n \approx 1$.}
\label{figure3}
\end{center}
\end{figure*}

Together the behaviour of $\omega_{K}$ we display, for all the components (see Fig.\ref{figure2}), the energy density value $\rho_{K}$ and the overall sum of these quantities\footnote{In order to achieve these plots we settle the quantity $N_0^2t^2=1$.}. It is evident that energy conditions are violated for certain values of $n$. For example it is easy to observe that the Weak Energy Condition, $\rho_{i} \ge 0$, for the first solution, is not satisfied by $\rho_{IMT}$ and alternatively by the other two energy densities $\rho_{Curv}$ and $\rho_{Mix}$.  For the second solution we have again no $n$ intervals where all components satisfy this energy prescription, since $\rho_{IMT}$ is quite always well behaved but $\rho_{Curv}$ and $\rho_{Mix}$ are alternatively negative. Finally, for the third solution we have a small region around $n \approx 1.5$ where all energy densities are positive, while the WEC is violated elsewhere. Although single matter-energy sources can violate the energy conditions, the total geometrized matter satisfies the WEC (dotted lines in each panel of Fig.\ref{figure2}) for all the ranges of the  parameter space in which cylinder solutions are defined. In addition, it has been recently showed that cosmological models with negative energy components can be suitably matched with observations if these ``disturbing" quantities are coupled with greater magnitude of positive  energy forms \cite{nemiroff2015}. Therefore the total top down geometrized matter is physical and preserves conceptions about energy conditions in 4-D. Such a result is quite customary when dealing with Induced Matter Theory models \cite{KK}. 
\\
Finally, let us discuss the cosmological evolution of the four-dimensional sections determined by each of solution (\ref{solcyl1})-(\ref{solcyl3}) and the  behaviour of the extra dimension.  Figure \ref{figure3} allow to draw significant informations about this behaviour describing the dependence of $a(t,\,y)$ and $\Phi(t,\,y)$ with respect to $n$ in the different cases. 
\\
In order to display an accelerating dynamics, a power law scale factor requires that the power index has to be negative or greater than one. In particular, universes expanding at an accelerating rate need $\alpha >1$ if we consider the definition given in Eqs.(\ref{plawpotenziali}).

\begin{figure*}[htbp]
\begin{center}
\includegraphics[width=12cm]{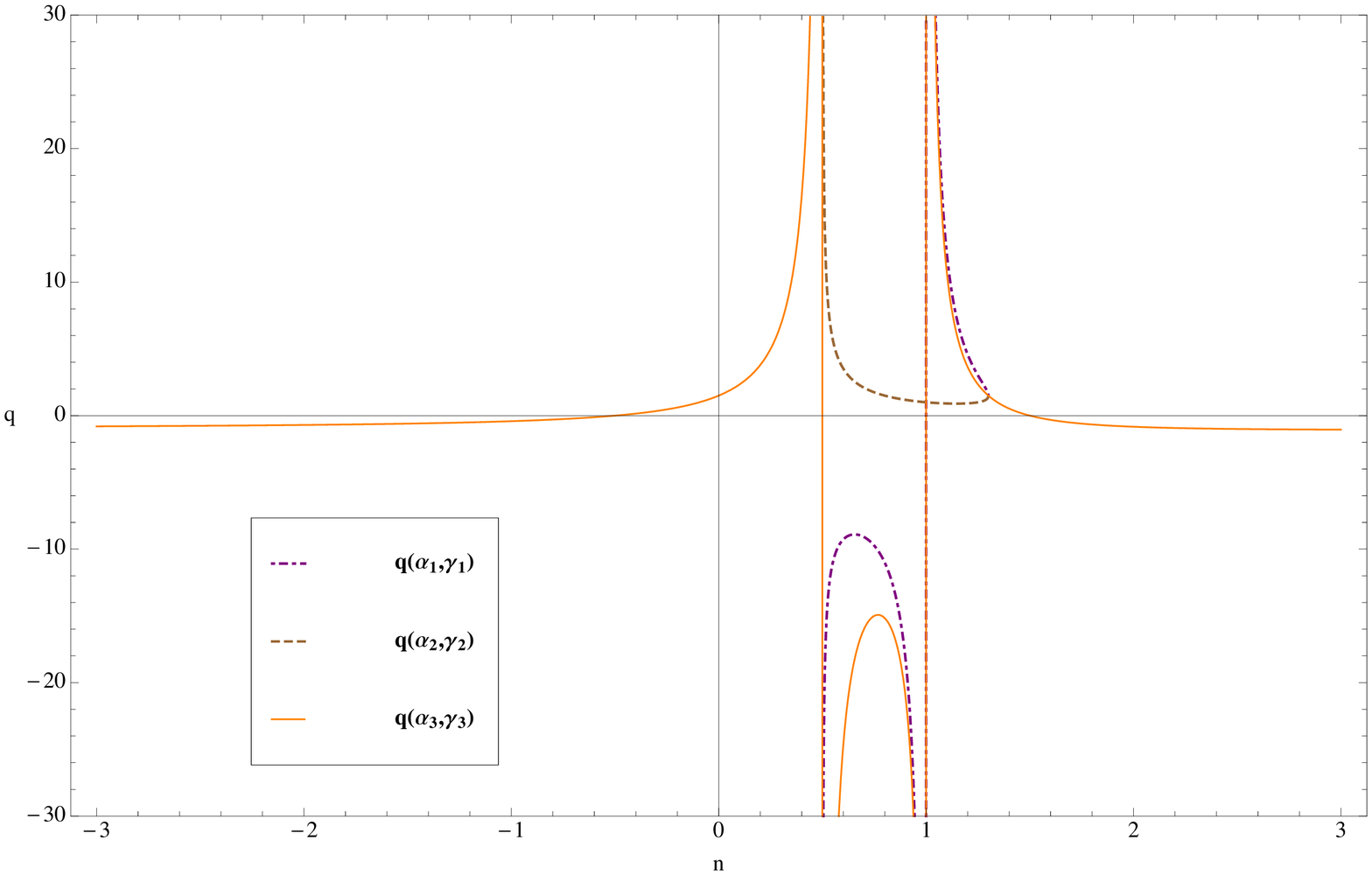}
\includegraphics[width=12cm]{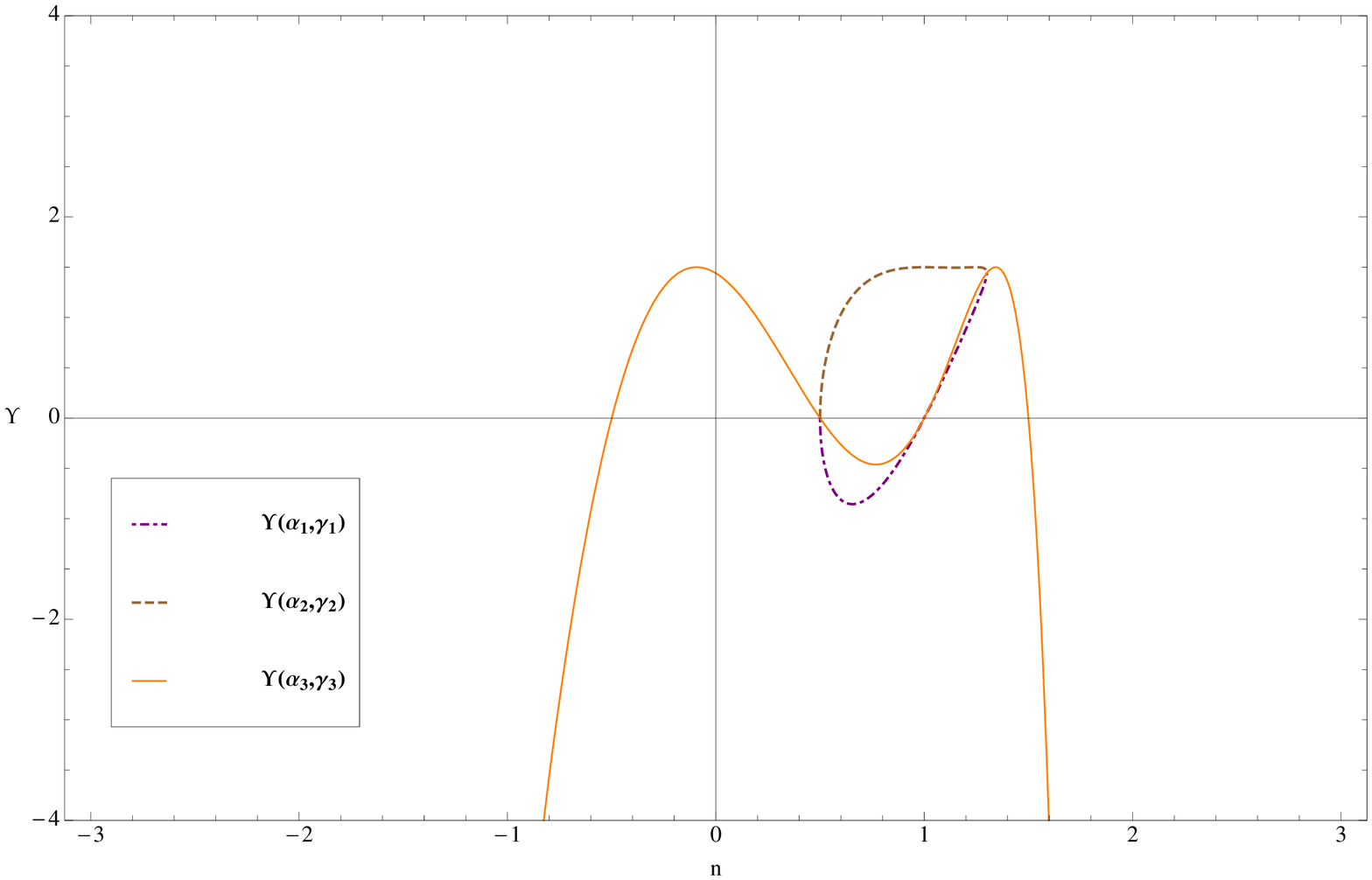}
\caption{The deceleration parameter and the $\Upsilon$ term, which enters the scale factor acceleration equation, vs. n. Again the three plot are related to the  cylinder solutions (\ref{solcyl1})-(\ref{solcyl3}).}
\label{figure4}
\end{center}
\end{figure*}

Actually, since our approach in the 4-D limit determines a multi-fluid model, the acceleration condition for the cosmological dynamics turns out to be characterized by the general relation for the scale factor  
$$\frac{\ddot{a}}{a}=-\frac{4\pi G}{3}(\rho_{tot}+3 p_{tot}).$$
Therefore, barotropic fluids like ones outlined by our model, with $\rho_{tot}=\sum_K\rho_K$ and $p_{tot}=\sum_K p_K$, imply that such a relation can be written as 
\begin{equation}
\frac{\ddot{a}}{a}=-\frac{4\pi G}{3}\sum_K\left(1+3\omega_K\right)\rho_K,
\end{equation}
and the condition for accelerating cosmologies becomes $\Upsilon=\sum_K\left(1+3\omega_K\right)\rho_K < 0$. As one can see in Fig. \ref{figure4} for our model the regions of the parameters space that fulfill this requirement exactly determine accelerating expansions (the deceleration parameter $q=-\frac{\ddot{a}a}{\dot{a}^2}$ is negative). Let us remember that the observations suggest $q \approx - 0.5$ \cite{giostri2012,waga2016}, albeit models considering power law solutions can also allow for higher limits i.e. $q \approx - 0.3$ \cite{rani2015}. 
\\
A peculiar aspect of our model is represented by the possibility that singles energy densities can achieve negative defined values. Of course, as said above, the overall evolution is still physical since the effective energy density budget is always positive definite. This behaviour, however, can affect the effective cosmological evolution. For example, it can prevent the possibility to obtain models with speeding up evolution that accord with cosmological observations. In fact, it is possible to observe (cfr. Fig.\ref{figure1} and Fig.\ref{figure4}) that there are regions of the parameter space where the induced cosmological fluids all assume negative barotropic factors while cosmological dynamics is still decelerating ($q\geq0$). On the other side, this peculiarity can also determine suitable conditions to frame models that experience a cosmic speeding up.

It is evident that, at this stage, our results can be only considered as a toy model achievement. Power law solutions can only describe limited stages of universe evolution. More general analyses are required in order to understand the real predictions of the model. In particular, the possibility to have a coexistence of negative defined equations of state with both decelerating dynamics and accelerating ones can represent an intriguing perspective to deepen in the framework of more general cosmological solutions.
\\
Let us discuss in some detail the results contained in Fig.\ref{figure1}-\ref{figure4}. 

\begin{itemize}
\item{Solution ($\alpha_1,\,\gamma_1$)-(\ref{solcyl1});

 This solution is allowed in the interval $\frac{1}{2}<n <\frac{13}{10}$. Values of $n<1$ give three negative EoS fluids within the cosmic pie. The GR limit ($n=1$) implies a trivial behaviour with vanishing $\rho_{K}$. For $n\sim1.3$ we have two standard matter components and the cosmological constant like source driven by $T^{Mix}_{\mu\nu}$. In such a case $\rho_{IMT},\,\rho_{Curv}>0$ while $\rho_{Mix}<0$, the whole cosmic mass-energy budget is positively defined as already said above.
 \\
Looking to the cosmological dynamics one can notice that this solution implies a negative power law index for the scale factor within the interval $\frac{1}{2}<n<1$ while $0\leq \alpha < 1$ for $1<n<\frac{13}{10}$. According with Fig.\ref{figure3}-\ref{figure4} this behaviour indicates accelerated contracting universes in the first interval and standard matter-like dominated cosmologies in the second one. The extra-dimension is always decreasing since is $0<\gamma<1$. It is interesting to notice that the ordinary matter-like regime is achieved: in the interval $1<n<1.25$ with two of the three fluids that behave as sources with a negative defined EoS (the other one is a stiffed fluid with $\omega_{IMT}>1$) and in the range $1.25<n<\frac{13}{10}$ with two standard matter-like sources ($0<\omega_{IMT}<1$ and $\omega_{Curv}>1$) together the cosmological constant contribute. The divergence of the barotropic factor $\omega_{Curv}$ for $n=1.25$ is regulated by the correspondent vanishing of $\rho_{Curv}$.}

\item{Solution ($\alpha_2,\,\gamma_2$)-(\ref{solcyl2});

We have again $\frac{1}{2}<n <\frac{13}{10}$. For this solution, the scale factor shows a standard matter rate since is always $0<\alpha<\frac{1}{2}$, there is no acceleration ($q({\alpha_2,\,\gamma_2})>0$).  The induced matter component $\rho_{IMT}$ is always a well behaved thermodynamical fluid with $0\leq\omega_{IMT}\leq 2/3$. The curvature induced sources can have a negative EoS or a positive one with $-5/6\leq\omega_{Curv}\leq 3/2$. For values $n\gtrsim 1.5$ we have two standard matter components and a cosmological constant like source.  Energy densities can be positive either negative defined. Around $n\sim 1.2$ we have two standard matter-like sources with positive $\rho_{i}$, while $\rho_{Mix}$ is negative, $\rho_{tot}$  is always $\geq 0$. For $n=1$, the two curvature induced sources collapse to a zero-valued energy density, while the induced matter tensor plays the role of a radiation fluid with $\omega_{IMT}=1/3$ as in usual KK-GR cylinder models. The fifth dimension can experience a reduction when $\frac{1}{2}<n<\frac{5}{4}$ and an increasing evolution for $\frac{5}{4}<n<\frac{13}{10}$, is static when $\gamma=\frac{1}{2},\,\frac{5}{4}$. }

\item{Solution ($\alpha_3,\,\gamma_3$)-(\ref{solcyl3});

Admissible values for $n\in{\mathbb{R}}-\{5/2\}$.  This solution explicitly resembles the standard 4-D $f(R)$-gravity solution given in \cite{curv2}. The scale factor and the extra-dimension show the same dynamics since $\alpha_3=\gamma_3$.  There are two interesting regions around $n\sim 0.2$ and $n\sim 1.6$. In the first case we have two standard matter components, dust-like and radiation-like, with $\omega_{Curv}\approx 0, \omega_{IMT}\approx 0.26$ together a cosmological constant-like one with $\omega_{Mix}=-1$. The correspondent scale factor is however non accelerating since $q$ is positive. The second region shows a dust-like source with $\omega_{Curv}\approx 0$ and two dark energy-like components, one of them phantom-like with $\omega_{IMT}\approx -2.6$. In such a case the 4-D section shows an accelerating expansion since $q<0$ ($q_0 = -0.5$ for $n \approx 1.69$) as well as an increasing extra-dimension. The zeros of $\omega_{IMT}$ (dust-like behaviour of this source) are trivial since the correspondent value of the energy density is vanishing. 
\\
Actually, this solution allows a plethora of different behaviours. In particular, expanding solutions with no acceleration are obtained for $-\frac{1}{2}\leq n < \frac{1}{2}$ and $1<n\leq \frac{3}{2}$ with two standard matter-like sources plus the effective cosmological constant contribute given by $T^{Mix}_{\mu\nu}$; on the other side accelerated expansions are given by models with $n<-\frac{1}{2}$  and $\frac{3}{2}<n<\frac{5}{2}$. In these intervals $T^{Curv}_{\mu\nu}$ plays the role of a baryonic component and the other two matter-energy sources act as dark energy-like sources. The energy densities have, again, both negative and positive values with $\rho_{tot} \geq 0$. For $n> 1.5$ we have all positive $\rho_{K}$ with $\omega_{K}\sim -1$, in such a case the three induced components behave all as cosmological constant-like sources. }

\end{itemize}

%. The induced fluid, in fact, in this case, consists only of photons \cite{KK}

As a matter of fact, fourth order gravity provide, in the cylindricity case, a completely different framework with respect to the standard Induced-Matter Theory. The standard scheme implies, in the 4-D limit, that induced matter behaves as a radiation-like fluid ($p=\rho/3$) \cite{KK} according with the result obtained with an initial condition of cylindricity by Kaluza \cite{kaluza1921}. In the case of higher order Kaluza-Klein models the energy-momentum tensors entering the 4-D induced field equations can describe more general  matter sources. Therefore one can obtain dynamically significant cosmologies also in presence of a Killing symmetry for the extra-coordinate. Of course, the significance of such solutions have to be corroborated  with cosmological and astrophysical data.

%%%%%%%%%%%%%%%%%%%%%%%%%%%%%%%%%%%%%%%%%%%%%%%%%%%%
\section{Conclusions}
%%%%%%%%%%%%%%%%%%%%%%%%%%%%%%%%%%%%%%%%%%%%%%%%%%%%

We discussed the possibility to develop a higher order and higher dimensional gravity model. Within such an approach, after the 5-D to 4-D reduction procedure, it has been possible to suggest a new interpretive scheme for the 4-D cosmological phenomenology.  Specifically, we have considered a vacuum 5-D $f(R)$-gravity model. The 4-D effective reduction provides, in this case, a GR-like cosmology characterized by means of three induced matter-energy tensors of geometrical origin. Within this   matter geometrization paradigm the cosmological fluid sources can be achieved considering: the standard Induced Matter Tensor of non-compactified KK theory $T^{IMT}_{\mu\nu}$, the Curvature Quintessence tensor $T^{Curv}_{\mu\nu}$ coming from higher order derivative terms of $f(R)$-gravity and $T^{Mix}_{\mu\nu}$ a mixed quantity that arises from both higher dimensional and higher order curvature counter-terms. This last quantity embodies the 5-D to 4-D reduction of the two combined approaches. We have obtained the complete field equations formalism and in order to check model prediction we have worked out a routine to search for power law solutions in the case $f(R)=f_0R^n$.  
The whole scheme provides the right KK-GR limit once $f(R)\rightarrow R$. In such a case, on shell, radiation dominated cosmologies ($p=1/3\rho$) are obtained if no dependence on the extra-coordinate is taken into account. More generally, the KK-GR limit furnishes solutions that agree with similar results already present in literature. In particular, the curvature induced matter-energy tensors vanish while the Induced Matter Tensor, that collects all metric terms depending on extra-coordinate, turns out to be the only one source of 4-D field equation. 
\\
Relaxing the hypotheses on $n$, i.e. by assuming a generic power law fourth order gravity model, it is possible to search for more general solutions. As a first preliminary step, cylindricity has been taken into account. A more general study has been left to a dedicated forthcoming publication. Differently from the standard KK-GR theory, the combination of fourth order gravity with higher dimensions implies, in such a case, non trivial solutions that can give rise to interesting cosmologies. We have obtained a set of solutions, parameterized by the $f(R)$ power index $n$, that determine a variety of possible barotropic fluids in the 4-D mass-energy sector. In relation to the value of $n$ one can have both standard matter components and dark energy-like ones. The cylinder condition implies that $T^{Mix}_{\mu\nu}$, according to its definition, behaves as an effective cosmological constant-like term with $\omega_{Mix}=-1$. Actually, in such a case, the energy density $\rho_{Mix}$ depends on time with an inverse square law. This fact suggests the intriguing possibility to have a time varying cosmological constant. The cosmological behaviour of the 4-D sections shows that it is possible to have both accelerating and non accelerating expansions. In particular, non accelerated expansion can be obtained also in presence of subdominant dark-energy like components. On the other side, it can be easily framed a dark energy dominated speeding up universe generalizing results obtained in the standard 4-D $f(R)$-gravity. The interesting aspect of this last solutions is that geometry can provide both standard matter-like sources and dark-energy-like ones, one of this can mimic a phantom fluid. The extra-dimension follows an independent evolution, that, in the third solution, is equivalent to one of the scale factor. Therefore, in such a case, expanding accelerated evolutions are characterized also by a growing fifth coordinate.
\\
The price to pay for geometrizing matter-energy sources is that cosmological energy densities can assume negative values violating the WEC. We have shown that although this is true for some values of $n$, also in interesting regions of the parameters space, the whole mass-energy budget $\rho_{tot}$ is always positive definite and, therefore, total top-down geometrized matter is physical. 

The fact that geometrically induced matter-energy tensors could, alternatively, play the role of standard matter components and of dark energy-like sources can represent an intriguing perspective in order to interpret cosmological observations. In particular, this achievement seems to deserve some more studies in the direction of dark energy and dark matter interpretation.  Of course, more insights are necessary in order to evaluate such a theoretical scheme. In particular, the matter sector (i.e. matter-like components), which has been satisfactory checked within standard IMT \cite{liu9298,billyard2001}, require a careful investigation since matter properties  should be completely induced by gravity.  In addition, model predictions have also to be checked also against cosmological and astrophysical data. 

Furthermore, if $f(R)$-Kaluza-Klein gravity determine, in the 5-D to 4-D reduction of cylinder solutions, cosmological models with a suitable top-down matter geometrization, in principle, more general solutions could allow to develop cosmological models with a wider phenomenology. On the other side, the fifth dimension role requires a careful analysis when the extra-coordinate Killing symmetry is discarded. In fact, if the cylinder condition is relaxed one naturally gets into the game of a length scale related with the extra-coordinate. In this sense, it has been shown that 5-D $f(R)$ gravity suggests that, eventually, there is a strict interconnection between the fifth dimension dynamics and the higher derivative counter-terms. 

%%%%%%%%%%%%%%%%%%%%%%%%%%%%%%%%%%%%%%%%%%%%%%%%%%%%

%
%%%%%%%%%%%%%%%%%%%%%%%%%%%%%%%%%%%%%%%%%%%
\vspace{0.3cm}
{{\bf Acknowledgments}: The author is thankful with Prof. S. De Pasquale and Prof. S. Pace for their help and support. } 
%
%
%%%%%%%%%%%%%%%%%%%%%%%%%%%%%%%%%%%%%%%%%%%%%


\begin{thebibliography}{99}
%%%%%%%%%%%%%%%%%%%%%%%%%%%%%%%%%%%%%%%%%%%%%


\bibitem{RiessPerlmutter}
A. G. Riess, {\it et al.}, AJ, {\bf 116}, 1009, (1998); S. Perlmutter, {\it et al.}, ApJ, {\bf 517}, 565, (1999).

\bibitem{SNeIa-2}
P. M. Garnavich, {\it et al.}, ApJ, {\bf 74}, 509, (1998); R. Rebolo, {\it et al.}, Mon. Not. Roy. Astr. Soc., {\bf 353}, 747, (2004);
A. C. Pope {\it et al.}, Astrophys. J., {\bf 607}, 655, (2004).

\bibitem{SNeIa-3}
M. Tegmark, {\it et al.}, (SDSS Collaboration), Phys. Rev. D, {\bf 74}, 123507, (2006);
W. J. Percival, S. Cole, D. J. Eisenstein, R. C. Nichol, J. A. Peacock, A. C. Pope, A. S. Szalay, Mon. Not. Roy. Astr. Soc., {\bf 381}, 1053, (2007).

\bibitem{sahnipeebles}
V.~Sahni and A.~A.~Starobinsky, Int. J. Mod. Phys. D, {\bf 9}, 373, (2000); P. J. E. Peebles, B. Ratra, Rev. Mod. Phys., {\bf 75}, 559, (2003).


\bibitem{burgess2013}
C. P. Burgess, ArXiv[hep-th]:1309.4133, \emph{The Cosmological Constant Problem: Why it's hard to get Dark Energy from Micro-physics}, DOI:$10.1093/acprof:oso/9780198728856.003.0004$, (2013).

\bibitem{sami2006}
E.J. Copeland, M. Sami, S. Tsujikawa, Int. J. Mod. Phys. D {\bf 15}, 1753, (2006). 

\bibitem{DEreview}
J. Yoo, Y. Watanabe, Int. J. Mod. Phys. D {\bf 21}, 1230002 (2012);  A. Joyce, L. Lombriser, F. Schmidt, Ann. Rev. Nucl. Part. Sci. {\bf 66}, 95, (2016).

\bibitem{Brans1961} 
C. Brans, R. H. Dicke, Phys. Rev. {\bf 124}, 124, (1961).

\bibitem{Bartolo1999} 
N Bartolo, M Pietroni,  Physical Review D {\bf 61 (2)}, 023518, (1999).

\bibitem{FaraoniST} 
V. Faraoni, Cosmology in scalar-tensor gravity, Kluwer Academic, Dordrecht, (2004).

\bibitem{curv1} S. Capozziello, S. Carloni, A. Troisi, Res. Dev. Astron. Astrophys., {\bf 1}, 625, (2003); S. M. Carroll, V. Duvvuri, M. Trodden, M. S. Turner, Phys.Rev. D {\bf 70}, 043528, (2004).
\bibitem{CapozFar} S. Capozziello and V. Faraoni, Beyond Einstein Gravity, Springer, Dordrecht, (2011).
\bibitem{NojOd} S. Nojiri, S. D. Odintsov, Phys. Rept. {\bf 505},  59-144, (2011). 

\bibitem{Dvali2000} G. Dvali, G. Gabadadze, M. Porrati , Phys. Lett. B {\bf 485}, 208, (2000).

\bibitem{brax2003}
P. Brax, C. van de Bruck, Class. Quant. Grav.{\bf  20}, R201, (2003).
\bibitem{MaartensLR} R. Maartens, K. Koyama, Living Rev. Relat. {\bf 13}, 5,  (2010).
\bibitem{RS} L. Randall and R. Sundrum, Phys. Rev. Lett. {\bf 83} (1999) 3370; L. Randall and R. Sundrum, Phys. Rev. Lett. {\bf 83} (1999) 4690.

\bibitem{wessbook1} P. S. Wesson, Space-Time-Matter, World Scientific Singapore (1999).
\bibitem{KK} J. M. Overduin, P.S. Wesson, Phys. Rept. {\bf 283} (1997) 302.
\bibitem{wessbook2} P. S. Wesson, Five-Dimensional Physics, World Scientific Singapore (2006).

\bibitem{f(R)Review} T. P. Sotiriou, V. Faraoni, Rev. Mod. Phys. {\bf 82}, 451, (2010); T.~Clifton, P.~G.~Ferreira, A.~Padilla and C.~Skordis, Phys. Rept.,  {\bf 513}, 1, (2012).  

\bibitem{curvmimick} 
S. Capozziello, V. F. Cardone, A. Troisi, Phys. Rev. D {\bf 71}, 043503, (2005).

\bibitem{husaw} 
W. Hu, I. Sawicki, Phys. Rev. D {\bf 76}, 064004, (2007).

\bibitem{starobL} 
A.A. Starobinsky,  JETP Letters {\bf 86.3}, 157, (2007).

\bibitem{curvlow} 
S. Capozziello, A. Stabile, A. Troisi, Phys. Rev. D {\bf 76 (10)}, 10401, (2007).

\bibitem{stelle} 
K.S. Stelle, Gen. Rel. Grav. {\bf 9}, 353, (1978); H.J. Schmidt, Int. J. Geom. Methods Mod. Phys. {\bf 04}, 209 (2007).

\bibitem{defeliceRev} 
A. De Felice, S. Tsujikawa, Living Rev. Relativity {\bf 13},  (2010); T.P. Sotiriou, V. Faraoni, Rev. Mod. Phys. {\bf 82}, 451, (2010).

\bibitem{alcaniz2007} 
J. Santos, J.S. Alcaniz, M.J. Rebou{\c{c}}as, F.C. Carvalho, Phys. Rev. D {\bf 76}, 083513, (2007).

\bibitem{nonminRmatt} 
O. Bertolami, C. G. Bohmer, T. Harko and F. S. N. Lobo, Phys. Rev. D{\bf 75} (2007) 104016; V. Faraoni, Phys. Rev. D{\bf 76},  127501, (2007). 

\bibitem{fRT} 
T. Harko, F. S.N. Lobo, S. Nojiri, S. D. Odintsov, Phys.Rev. D {\bf 84}, 024020, (2011).

\bibitem{nordstrom1914}
G. Nordstr{\o}m, Phys. Z. {\bf 15}, 504 (1914).

\bibitem{kaluza1921}
T. Kaluza, Sitz. Preuss. Akad. Wiss. {\bf 33}, 966 (1921).
\bibitem{klein1926}
O. Klein, Z. Phys. {\bf 37}, 895 (1926).

\bibitem{aguilar2008}
J.E.M. Aguilar, C. Romero and A. Barros, Gen. Rel. Grav. 40, 117 (2008).
\bibitem{PdeLeon2010} 
J. Ponce de Leon, Class. Quant. Grav. {\bf 27}, 095002 (2010).
\bibitem{deLeonjcap2010} 
J. Ponce de Leon, JCAP {\bf 03}, 030 (2010).
\bibitem{rasoulietal} 
S.M. M. Rasouli, M. Farhoudi and H.R. Sepangi, Class. Quant. Grav. 28, 155004 (2011); A.F. Bahrehbakhsh, M. Farhoudi and H. Shojaie Gen. Rel. Grav. {\bf 43}, 847 (2011); A. F. Bahrehbakhsh, M. Farhoudi, and H. Vakili, Int. J. Mod. Phys. D {\bf 22}, 1350070 (2013).

\bibitem{curv2} 
S. Capozziello, V.F. Cardone, S. Carloni, A. Troisi, Int. J. Mod. Phys. D {\bf 12}, 1969, (2003); 

\bibitem{curv3}
S. Capozziello, V.F. Cardone, A. Troisi, J. Cosmol. Astropart. Phys., {\bf (08)}, 001, (2006).


\bibitem{schimming2002}
R. Schimming, M. Abdel-Megied, F. Ibrahim, Chaos, Solitons and Fractals {\bf 14-8},  1255, (2002).

\bibitem{schimming2003}
R. Schimming, M. Abdel-Megied, F. Ibrahim, Chaos, Solitons and Fractals {\bf 15-1},  57, (2003).

\bibitem{vajdi2009}
A. Aghmohammadi, Kh. Saaidi, M.R. Abolhassani, A. Vajdi, Phys. Scripta {\bf 80}, 065008, (2009).

\bibitem{huang2010}
B. Huang, Song Li, Y. Ma, Phys. Rev. D {\bf 81}, 064003, (2010).

\bibitem{wu2014}
Y.B. Wu et al, Eur. Phys. J. C {\bf 74}, 2791, (2014).

\bibitem{darabi2009}
  K.~Atazadeh, F.~Darabi and H.~R.~Sepangi,
  %``Correspondence between modified gravity and 5D Ricci-flat cosmologies,''
  Int.\ J.\ Mod.\ Phys.\ D {\bf 18}, 1049, (2009).

\bibitem{aguilar2015} 
  J.~E.~M.~Aguilar,
  %``New effective coupled $F(^{(4)}\!R,\varphi )$ modified gravity from $f(^{(5)}\!R)$ gravity in five dimensions,''
  Eur.\ Phys.\ J.\ C {\bf 75}, no.12,  575 (2015).


\bibitem{buchdal}
H.A. Buchdahl,  Monthly Notices of the Royal Astronomical Society {\bf 150}, 1, (1970).

\bibitem{oneloop}
R. Utiyama et al., J.Math.Phys. {\bf 3}, 608, (1962); 
I.L. Buchbinder, S.D. Odintsov, I.L. Shapiro., Effective action in quantum gravity - IOP. Bristol, (1992).

\bibitem{starobinsky}
A.A. Starobinsky, Phys. Lett. B {\bf 91}, 99, (1980). 

\bibitem{kerner} 
R. Kerner, Gen. Rel. Grav. {\bf 14}, 453, (1982).

\bibitem{woodard}
R. P. Woodard, Lect. Notes. Phys. {\bf 720},  403, (2007);  R. P. Woodard, arXiv:1506.02210 [hep-th].

\bibitem{carlip}
S. Carlip,  Reports on progress in physics {\bf 64.8}, 885, (2001).

\bibitem{kaku} M. Kaku, Introduction to Superstrings and M-Theory, Springer, (1999).


\bibitem{suNgra} Y. Choquet-Bruhat, ``Supergravities and Kaluza-Klein theories." Topological properties and global structure of space-time. Springer US, 31-48, (1986); O. Aharony et al., Physics Reports {\bf 323.3}, 183, (2000).

\bibitem{cambell1926} J.E. Campbell, A course of Differential Geometry, Clarendon Press, Oxford, (1926).

\bibitem{magaard1963} L. Magaard, ``Zur Einbettung Riemannscher R$\Ddot{a}$ume in Einstein-R$\Ddot{a}$ume und konform-euklidische R$\Ddot{a}$ume?, PhD Thesis, University of Kiel, (1963).

\bibitem{romero96} 
C., Romero, R. Tavakol, R. Zalaletdinov,  Gen. Rel. Grav.,  {\bf 28.3}, 365, (1996).

\bibitem{wesson1992} P.S. Wesson, Astrophys. J., {\bf 397}, L91, (1992).


\bibitem{kernerEmb}
R. Kerner, Gen. Rel. Grav. {\bf 9 (3)}, 257, (1978); R. Kerner, S. Vitale, Int. J. Mod. Phys. A {\bf 24}, 1465 (2009).


\bibitem{wessPdL92}
P.S. Wesson and J. Ponce de Leon, 
%``Kaluza-Klein equations, Einstein's equations, and an effective energy-momentum tensor", 
{J. Math. Phys.} {\bf 33}, 3883 (1992).

\bibitem{PdLMPLA01} J. Ponce de Leon, Mod. Phys. Lett. A {\bf 16}, 2291, (2001).

\bibitem{cmb} C.L. Bennett, et al., Astrophys. J. Letters {\bf 464}, L1, (1996); P. de Bernardis, et al., Nature {\bf 404} (6781), 955, (2000); G. Hinshaw,  et al., Astrophys. J. Suppl. {\bf 170}, 288, (2007).

\bibitem{planck} Planck Collaboration: P. A. R. Ade, et al., arXiv:1502.01589; Planck Collaboration: P. A. R. Ade, et al., arXiv:1502.01593.

\bibitem{curvgalaxies} S. Capozziello, V.F. Cardone, A. Troisi, Mon. Not. Roy. Astron. Soc. {\bf 375}, 1423, (2007).

\bibitem{WessonP} P.S. Wesson, J. Ponce de Leon,  Gen. Rel. Grav. {\bf 26}, 555, (1994).

\bibitem{PdL88} J. Ponce de Leon, %?Cosmological models in a Kaluza-Klein theory with variable rest mass?, 
Gen. Rel. Grav. {\bf 20}, 539 (1988).


\bibitem{lopez1996} 
  J.~L.~Lopez and D.~V.~Nanopoulos,
  %``A New cosmological constant model,''
  Mod.\ Phys.\ Lett.\ A {\bf 11}, 1, (1996)

\bibitem{overcoop1998}
  J.~M.~Overduin and F.~I.~Cooperstock,
  %``Evolution of the scale factor with a variable cosmological term,''
  Phys.\ Rev.\ D {\bf 58}, 043506, (1998) .
 
 
 \bibitem{liu9298} 
  H.~Y.~Liu and P.~S.~Wesson,
  %``Exact solutions of general relativity derived from 5-D 'black hole' solutions of Kaluza-Klein theory,''
  J.\ Math.\ Phys.\  {\bf 33}, 3888, (1992); 
  P.~S.~Wesson and H.~Y.~Liu,
  %``Shell-like solutions in Kaluza-Klein theory,''
  Phys.\ Lett.\ B {\bf 432}, 266 (1998).

\bibitem{socorro1996} 
  J.~Socorro, V.~M.~Villanueva and L.~O.~Pimentel,
  %``Classical solutions in five-dimensional induced matter theory and its relation to an imperfect fluid,''
  Int.\ J.\ Mod.\ Phys.\ A {\bf 11}, 5495, (1996).
  

\bibitem{nemiroff2015} 
R.J. Nemiroff, R. Joshi, B.R. Patla, Journ. of Cosmol. Astrop. Physics (2015).


\bibitem{giostri2012} 
R. Giostri, M. Vargas dos Santos, I. Waga, et al., JCAP {\bf 03}, 027, (2012).

\bibitem{waga2016}  
M. Vargas dos Santosa, R. R. R. Reisa, I. Waga, JCAP {\bf 02}, 066, (2016).

\bibitem{rani2015} 
S. Rani, A. Altaibayeva, M. Shahalam, J.K. Singha,  R. Myrzakulov, JCAP {\bf 03}, 031, (2015).

\bibitem{billyard2001} 
  A.~P.~Billyard and W.~N.~Sajko,
  %``Induced matter and particle motion in noncompact Kaluza-Klein gravity,''
  Gen.\ Rel.\ Grav.\  {\bf 33}, 1929, (2001).



\end{thebibliography}
\end{document}